\documentclass[aps,prb,twocolumn,eqsecnum]{revtex4-1}

\pdfoutput=1

\usepackage[T1]{fontenc}
\usepackage[latin9]{inputenc}
\usepackage{amsmath}
\usepackage{graphicx}
\usepackage{esint}
\usepackage{longtable}

\begin{document}
\title{Competing Nematic, Anti-ferromagnetic and Spin-flux orders in the Ground State of Bilayer Graphene}

\author{Y. Lemonik}
\affiliation{Physics Department, Columbia University, New York NY 10027, USA}

\author{I. Aleiner}
\affiliation{Physics Department, Columbia University, New York NY 10027, USA}

\author{V. I. Fal'ko}
\affiliation{Physics Department, Lancaster University, Lancaster, LA1 4YB, UK}

%\date{\today}
\begin{abstract}
  We analyze the phase diagram of the Bilayer graphene (BLG) at zero
  temperature and doping. Assuming that at the high energies the
  electronic system of BLG can be described within a weak coupling
  theory (consistent with the experimental evidence), we
  systematically study the evolution of the couplings with going from
  high to low energies.  The divergences of the couplings at some
  energies indicates the tendency towards certain symmetry
  breakings. Carrying out this program, we found that 
the phase diagram is determined by 
  microscopic couplings defined on the short distances (initial
  conditions).  We explored all plausible space of these initial
  conditions and found that the three states have the largest phase
  volume of the initial couplings: nematic, antiferromagnetic and spin
  flux (a.k.a quantum spin Hall). In addition, ferroelectric and
  two  superconducting phases and appear only
  near the very limits of the applicability of the weak coupling
  approach.

The paper also contains the derivation and analysis of the
renormalization group equations and the group theory classification of
all the possible phases which might arise from the symmetry breakings
of the lattice, spin rotation, and gauge symmetries of graphene.

\end{abstract}
\maketitle

\section{Introduction}

\begin{figure}
\includegraphics[width=1\columnwidth]{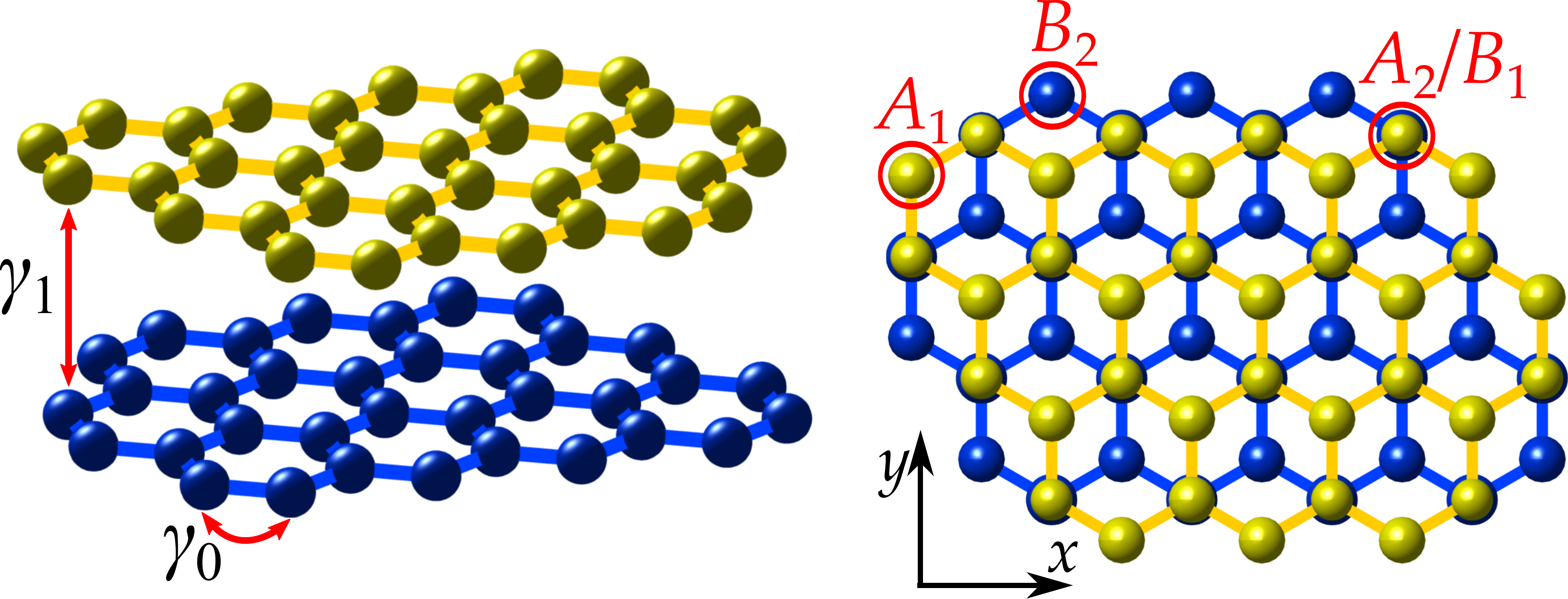}

\caption{ (Color online)Left panel: 3D view of bilayer graphene. 
The sites that sit on top of each other, connected by dotted lines, hybridize strongly and form bands with a gap of $\gamma_1 \approx 0.4 eV$. 
The low energy electron live on the half of the carbon atoms that sit over/under the centers of the hexagons. Right panel:
top-down view of the lattice.}
\label{fig:BLG3D}
\end{figure}

Bilayer graphene\cite{McCann,Novo1} (BLG) is a crystal which consists
of two monolayers of honeycomb carbon lattice arranged according the
Bernal stacking known from bulk graphite \cite{Novo1}.  In a Bernal
stacked lattice, one out of the two sites on the upper monolayer
resides directly over a site on the lower lattice, and the the other
carbon atoms are on/under the centers  of the hexagons (see
Fig.~\ref{fig:BLG3D} ).  Such a crystal has a very high symmetry with
symmetry group $\mathcal{D}_{3d}$.

This high symmetry may be lifted by the formation of correlated states
of electrons.  There is a plethora of ways the symmetry can be lifted,
some of which have been discussed in the recent literature: the
ferroelectric-layer asymmetric state \cite{Levitov,MacDonald4}, the
layer polarized antiferromagnetic
state\cite{Kharitonov,MacDonald2,Vafek2}, the quantum anomalous Hall
state \cite{Levitov2,MacDonald4,MacDonald3}, the "spin flux"/ quantum
spin Hall state\cite{MacDonald4,MacDonald3}, the charge density wave
state\cite{Dahal} and an anisotropic nematic liquid\cite{Letter,Vafek}.  
Some of the proposed phases above have a gap in the
electronic spectrum (ferroelectric, antiferromagnetic,
spin-flux, CDW), whereas in the other phases (nematic, ferromagnetic)
no gap is formed.  This large variety of possibilities makes the
theory of electronic properties of BLG a very interesting and
challenging subject.  The complexity of the theoretical problem is
compounded by two factors.  One is a lack of precise information about
the relevant interaction constants which determine the electronic
phase in undoped pristine BLG.  The other issue is the competition
between exchange energy contributions for a large number of candidate
phases which makes the determination of the ground state non-trivial, even
with precise knowledge of the interaction constants.

On the experimental side, several contradicting observations have been reported based on interpretations of the measured transport properties 
of suspended samples in terms of a gapful or gapless spectrum of electronic excitations\cite{Yacoby,Weitz,Mayorov, Velasco}. 
However, all of these works 
as well as optical studies of BLG \cite{Rotenberg,Henriksen,Fogler,Henriksen2,Kuzmenko1,Kuzmenko2}
indicate that the high-energy properties (but below $0.2 eV$) of BLG are well described by the two band model \cite{McCann} without interactions.  
This makes a comprehensive theoretical treatment of the problem starting from the weak coupling even more timely. 
In this paper, we employ the previously developed RG approach \cite{Letter} to identify the possible scenarios of symmetry breaking phase transition in BLG at low temperature and zero carrier density.

The tendency to form a state with spontaneously broken symmetry is encoded in the system response to local symmetry breaking fluctuations, in particular in their mutual interaction,
\begin{equation}
H_{int} \sim \int d^2\,r\sum_{\mathcal{A}\in IrReps} g_\mathcal{A} \delta\hat{\rho}_{\mathcal{A}}(r)\cdot\delta\hat{\rho}_{\mathcal{A}}(r).
\label{eq:HIntIntro}
\end{equation}
Here $\delta\hat{\rho}_\mathcal{A}$ are operators creating local density fluctuations breaking lattice symmetry, with $\delta\hat{\rho}_\mathcal{A} = \psi^\dagger\hat{M}\psi$  expressed in terms of electron annihilation and creation operators $\psi$ and $\psi^\dagger$, and $g_\mathcal{A}$ are coupling constants.   Each of the fluctuations $\delta\hat{\rho}_\mathcal{A}$ belong to one of the irreducible representations $\mathcal{A}$  (IrReps) of the symmetry group of the lattice.
(A precise definition of the densities can be found in Sec.~\ref{SEC2}).

If Hamiltonian (\ref{eq:HIntIntro}) is dominated by one term with
negative constant $g_\mathcal{A}$, we would expect it to be
energetically favorable for a state with a non-zero expectation value
of $\delta\hat{\rho}_\mathcal{A}$ to form, with the symmetry of the
ground state determine by the corresponding IrRep, $\mathcal{A}$.
However if the coupling constant in the dominant term is positive,
then the ground state is determined by the exchange energy, which can
be negative for not only for magnetic (ferro/antiferro) but also for non-magnetic
orderings, because of the sublattice/valley matrix structure. Because
of the large number of IrReps, this can result in a competition between
many phases. Therefore, to determine the ground state of BLG we must
know all the interaction constants $g_\mathcal{A}$ sufficiently well,
especially when the dominant ones are positive. The situation is
actually even more intriguing since attraction may result in a
superconducting phase with non-trivial Cooper pair structure.

\begin{figure*}
\includegraphics[width=1.6\columnwidth]{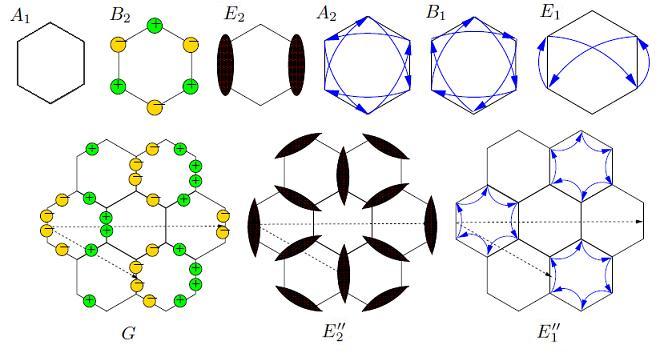}
\caption{ (Color online)Sketches  of the density and currents transforming according representations of the group ${\cal D}_{3d}^{\prime\prime}$. 
In the case of a spin singlet symmetry breaking, the plus and minus signs represent charges, the blue lines persistent currents and the black bars represent bonds. 
The $G$, $E_2$ and $E_2''$ order parameters triple the unit cell; the new Bravais lattice vectors are given by the dashed arrows. The representations are given in terms of Pauli matrices in Eqs.~(\ref{eq:gIrreps}) and (\ref{eq:DefHSR}). }\label{fig:Irreps}\end{figure*}

To add to the complexity of the problem, the values of the "constants" $g_\mathcal{A}$ are not fixed. They change as a function of the energy scale $\mathcal{E}$ within which the electrons establish the symmetry breaking correlations. 
The energy scale dependence, $g_{\mathcal{A}}(\mathcal{E})$, may be calculated using the renormalization group (RG) approach.
 In the RG approach the highest energy electron states are eliminated and their effects incorporated into a redefinition of the parameters of the theory. 
The renormalization of BLG parameters starts at the energy scale $\gamma_1/2 \approx 0.2eV$ which limits the applicability of the two-band model with parabolic spectrum and initial conditions $g_\mathcal{A}(\gamma_1/2)$. Then it is iterated until the lowest energy scale $\mathcal{E}$ is reached. 
This energy scale $\mathcal{E}$ is determined when the interaction energy in at least one of the the channels becomes of the order of  kinetic energy. After this scale is reached the mean field theory can be used to establish the electronic ground state.  
The necessary RG equations for the constants $g_{\mathcal{A}}$  and their interplay with Coulomb interaction,
\begin{equation}
H_C \sim \int d^2rd^2r' \frac{\psi^\dagger(r)\psi(r) \psi^\dagger(r')\psi(r')}{|r-r'|},
\label{eq:HCIntro}
\end{equation}
have been derived for the full set of eight constants in Refs. ~\onlinecite{Letter}. (Similar in spirit treatment of Ref.~\onlinecite{Vafek2} replaced Eq.~(\ref{eq:HCIntro})
with the short range weak interaction.)

Calculating  $g_\mathcal{A}(\gamma_1/2)$ requires detailed knowledge of the microscopic orbitals which is not available at present. Therefore, in this paper we explore a wide variety of initial conditions $g_\mathcal{A}(\gamma_1 /2)$  for the RG to find possible electronic ground states for BLG. We can make some arguments to constrain the values of the $g_\mathcal{A}(\gamma_1/2)$.  The coupling $g_{B_2}$ which describes the interaction of dipoles oriented perpendicular to the bilayer (see Fig. \ref{fig:Irreps}) must be positive at high energy scales.
 The four "current-current" interactions $g_{A_2}$, $g_{B_1}$, $g_{E_1}$  and $g_{E_1''}$ 
are only generated by virtual processes because of time reversal symmetry. 
Therefore we will set them to be zero at $\gamma_1/2$.

\begin{figure}
\includegraphics[width=0.9\columnwidth]{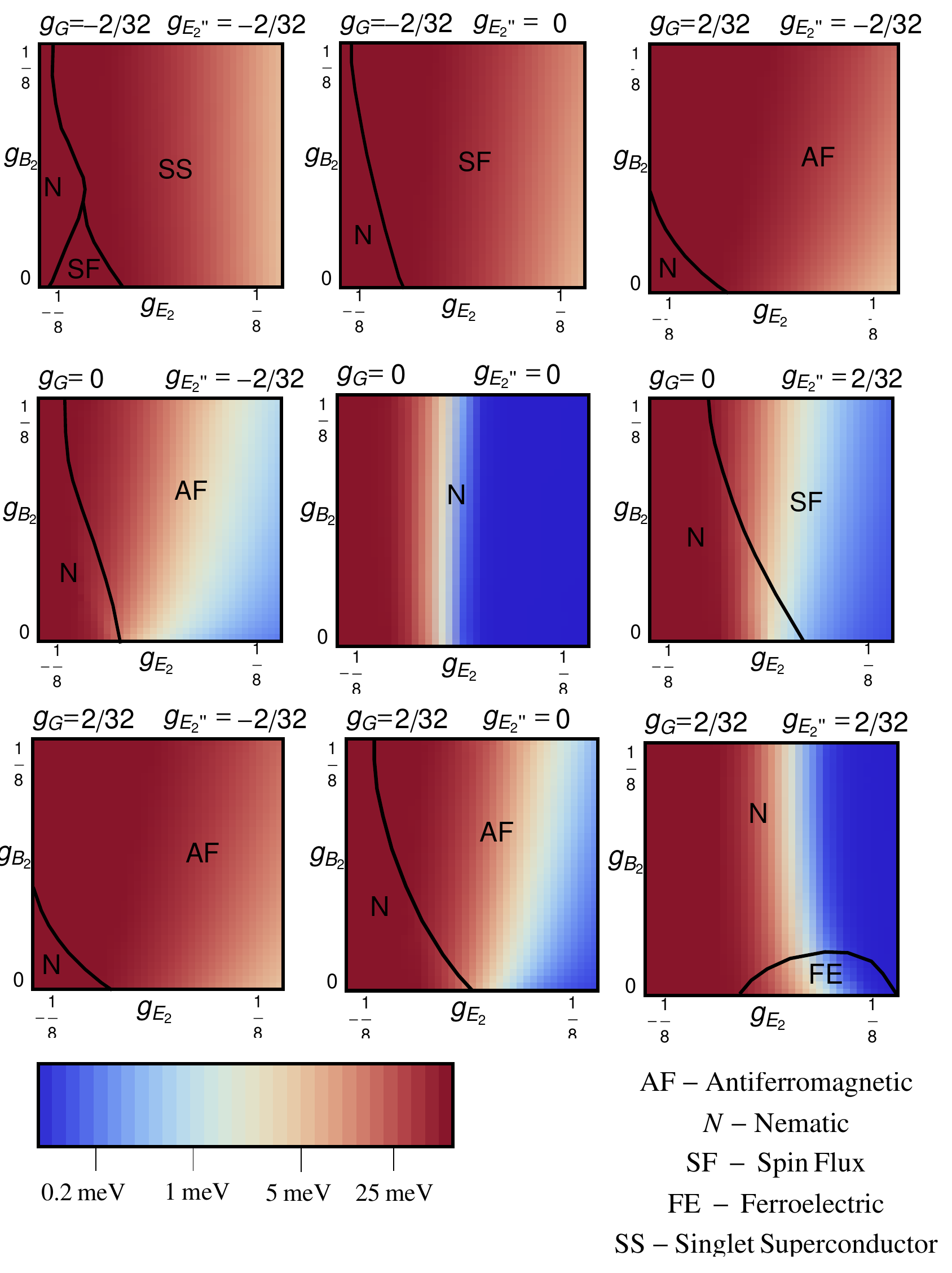}
\caption{ (Color online)Four cuts through the possible parameter space of BLG. The
  predicted gap (or saddle point energy for only gapless nematic phase) is indicated by the color scale and the predicted
  phase is indicated. N is a nematic, AF is an antiferromagnetic
  phase, SF is a spin flux phase and FE is a ferroelectric. A fifth
  predicted superconducting phase is in a range parameters not shown,
  see Fig.~\ref{fig:bigPlots} for more details. The $g$ are
  coupling constants of BLG, with the subscript labeling the
  irreducible representations in accordance with Fig. \ref{fig:Irreps}, and defined in Sec.~\ref{SEC2}.
All boundaries are the first order phase transitions.
}
\label{fig:OpenPlots}
\end{figure}

Also, it is interesting to note that  in the value of
$g_\mathcal{A}(\gamma_1/2)$ one has to take account of the
interactions between electrons via polarization of the
lattice. Particularly, the in-plane TO-LO phonons at the
$\Gamma$-point and TO phonons at the Brillouin zone corner have
energies comparable to $\gamma_1/2$, so that they mediate an
attractive interaction via their virtual creation/absorption. These
would give negative contributions to the bare values of $g_{E_2}$ and
$g_{E_2''}$.  
Analogously, virtual $LO-LA$ phonons from $K$ - the Brillouin zone corners give negative contribution
to the value of $g_G$.
Therefore, we make no assumption about the sign of
$g_{E_2}$, $g_{E_2''}$, and $g_G$.  A set of typical outcomes of the RG flow
and the resulting electronic phases is shown in
Fig. \ref{fig:OpenPlots}.

In Fig. \ref{fig:OpenPlots} we reproduce the earlier reported
result\cite{Letter,Vafek} that for the initial choice of
$g_\mathcal{A} = 0$ the RG flow leads to a nematic phase. The nematic
phase is a state with broken rotational (but intact translational)
symmetry corresponding to representation $E_2$ in
Fig. (\ref{fig:Irreps}), mimicking the effect of anisotropic hopping
along bonds with different directions on the honeycomb lattice. This
breaks the six-fold rotational symmetry by selecting one of axes of
the lattice. In this state the electronic spectrum remains gapless but
is significantly reconstructed from the unbroken symmetry state with
two four-fold degenerate Dirac cones at low energy. The state has the
same symmetry and spectrum as uniaxial strain\cite{BLGStrain}, and we
expect that strain will, all else equal, favor the nematic phase.
Figure \ref{fig:OpenPlots} shows that the nematic phase is the
preferred ground state not only when $g_\mathcal{A}(\gamma_1/2) = 0$,
but in a significant section of the $g_\mathcal{A}(\gamma_1/2)$
parameter space. In particular, the nematic phase always emerges from
the part of the parameter space where bare electron-electron couplings
causing intervalley scattering are zero ($g_G = g_ {E_2''} = g_{E_1''}
= 0$).

In other parts of the parameter space explored in this work  and illustrate in 
Fig. \ref{fig:OpenPlots}, the ground
state appears to be anti-ferromagnetic (AF), with the $A_1$ and $B_2$ sublattices of
  two layers, see Fig.~\ref{fig:BLG3D}, are spin
polarized in opposite directions. In the AF state the electronic
excitations are gapped (though neutral spin wave excitations are
gapless). Although the AF state prevails over a significant section of
the parameter space, the combinations of high energy couplings which
produce the AF state are not intuitive. For example, increasing the
bare coupling $g_{B_2}$ does not necessarily introduce the AF
phase. However increasing the bare coupling $g_G$ makes the ground
state AF. The reason for this counter-intuitive behavior is in the
complexity of the RG flows. Since there are eight non-linearly coupled
variables in the RG equations \cite{Letter, Vafek2}, the RG flow is
quite complicated, and the connection between the couplings at low
energy and the bare couplings at high energy is not obvious.

Exploring a broader parameter space further we find more phases. A
spin flux phase is found in a significant sector of the parameter
space $g_\mathcal{A}(\gamma_1/2)$, as seen in
Fig.~\ref{fig:OpenPlots}. This spin flux phase is a state with a
persistent spin current circling the honeycomb lattice rings,
corresponding to the spin triplet form of representation $B_1$ in
Fig.~\ref{fig:Irreps}. It may be viewed as the spontaneous formation
of a strong spin-orbit coupling. It therefore leads to a gapped
electronic spectrum and possibly a quantum spin Hall effect.

There are two more phases which appear to some degree in the phase
space explored. One is a ferroelectric phase (FE). The FE phase a trivial band gap insulator
where the bilayer becomes spontaneously charged like a capacitor. It
is a completely gapped phase. It corresponds to representation $B_2$
precisely the same representation as AF but spin singlet, rather than
spin triplet. Therefore, positive $g_{B_2}$ suppresses the
ferroelectric phase, which appears in Fig.~\ref{fig:OpenPlots} only in the fine tuned corners
corresponding to the applicability of the weak-coupling theory.

We also found other phase a new superconducting phase (not shown in figure,
see Fig.~\ref{fig:bigPlots} for more details) which has the energy 
tantalizingly close to the nematic and ferro-electric states.  It is a triplet
superconductor with a nontrivial Cooper pairing. Cooper pairs are
formed between pairs of electrons with opposite valleys and opposite
layer. The pairing is symmetric in exchange of valleys, but
antisymmetric in exchange of layers.

As usual, the singlet superconductivity appears only for the attractive interaction.
From the first panel on Fig.~\ref{fig:OpenPlots}, we see that it requires quite siginificant attraction
in two channels $g_{E_2''},g_G < 0$.

Below we describe how the conclusions listed above have been
reached. In Sec.~\ref{SEC2} we review the structure of BLG, it's
symmetry group and the low energy Hamiltonian. Section \ref{SEC3}
describes the resummation of the Coulomb interactions in the $1/N$
expansion\cite{Son,AKT,Son2,Letter}, where $N=4$ is the degeneracy of
the single particle spectrum. We then derive the RG equations that
connect the couplings at low and high energy scales.  In
Sec. ~\ref{SEC4} the results of the RG flow equations are analyzed and
augmented by a self-consistent mean field theory which produces a
possible phase.  Section \ref{SEC5} discusses the properties of the
emerging phases.  In Appendix we describe the group-theoretic analysis
of the phases of the BLG.  diagram.

\section{Model}
\label{SEC2}
The top view of the BLG lattice with Bernal stacking is shown on the right panel of Fig.~\ref{fig:BLG3D}. 
Here we label the two layers 1 and 2  and the four inequivalent lattice sites $A_1$, $ B_1$, $A_2$, $B_2$, with $A_2$ directly over $B_1$.

Calculation based on the minimal tight-binding model has established the following BLG band structure \cite{McCann}. The
$A_2$ and $B_1$ sites hybridize strongly and host states from the high energy bands with excitation energies $> \gamma_1 \approx 0.4eV$. The low energy fermionic excitations in BLG
belong to a four-component representation of the group $\mathcal{D}_{3d}$,
exactly as in monolayer graphene. The four fermionic fields $\psi$  are conveniently joined into a 4-vector as follows
\begin{equation}
 \vec{\psi}^{t}\equiv\left\{(\psi_{K}^{A},\psi_{K}^{B})^{AB},(\psi_{K'}^{B},-\psi_{K'}^{A})^{AB}\right\}_{KK'},
\label{eq:defvectorpsi}
\end{equation}
where the $\psi$s are true spinors including real electron spin.
 This four dimensional space can be written as the direct product of the (AB) and (KK') spaces. We will use this 
to write all operators as the sum of direct products $\tau^{AB}_{a}\tau^{KK'}_{b}\sigma_{c}$ of Pauli matrices in each space.
 We define $\{\tau_{i}^{AB},\tau_{i}^{KK'},\sigma_{i}\}$
as the Pauli matrices acting on layer, valley and spin, respectively,
and define  $\tau_{0}\equiv\openone$, $\tau_{\pm}\equiv(\tau_{x}\pm i\tau_{y})/2$.

The symmetries of the BLG lattice consist of the two independent lattice translation $\hat{t}_{1}$ and $\hat{t}_{2}$,
a $\hat{C}_{3}$, rotation by $2\pi/3$ around one of the lattice sites;
and two independent reflections: $\hat{R}_{h}$, reflection across
the y-axis, and $\hat{R}_{v}$, reflection across the x-axis together
with reflection through the plane midway between the graphene sheets, see right panel of Fig.~\ref{fig:BLG3D}. The reflections and rotations form
the point group $\mathcal{D}_{3d}$. The groups  $\mathcal{D}_{3d}$  and $\mathcal{C}_{6v}$ are isomorphic and have precisely 
the same action on the plane.  We also
ignore the spin-orbit interaction which gives an additional $SU(2)$ symmetry from the independent rotation of the spin. We
will be concerned with the physics about $K$ and $K'$ points which
are inequivalent in the Brillouin zone but are connected by $\hat{R}_{h}$. Rather
than dealing with two degenerate but inequivalent points we can triple
the unit cell, which maps $K$ and $K'$ onto the $\Gamma$ point.
In this view, the point group $D_{3d}$ is expanded to $\mathcal{D}_{3d}^{''}= \mathcal{D}_{3d}+\hat{t}_{1}\mathcal{D}_{3d}+\hat{t}_{2}\mathcal{D}_{3d}$ 
with the translation operator $\hat{t}_{1}$ with $\hat{t}_{1}^{2}=\hat{t}_{2}$
and $\hat{t}_{1}^{3}=\hat{1}$ (see {\em e.g.} Ref.~\onlinecite{Basko}).   

Ignoring the spin structure, the vector $\psi$ transforms as follows under the action of the symmetry
operators:

\begin{eqnarray}
\hat{t}_{1}\psi(\boldsymbol{r}) & = & e^{\frac{2\pi i}{3}\tau_{z}^{KK'}}\psi\left(t_1\boldsymbol{r}\right), \nonumber \\
\hat{C}_{3}\psi(\boldsymbol{r}) & = & e^{\frac{4\pi i}{3}\tau_{z}^{AB}}\psi\left(C_3\boldsymbol{r}\right), \nonumber\\
\hat{R}_{h}\psi(\boldsymbol{r}) & = & \tau_{y}^{AB}\tau_{y}^{KK'}\psi\left(R_h\boldsymbol{r}\right), \nonumber\\
\hat{R}_{v}\psi(\boldsymbol{r}) & = & \tau_{x}^{AB}\tau_{z}^{KK'}\psi\left(R_v\boldsymbol{r}\right), 
\end{eqnarray}
There is also the time reversal symmetry operation given by 
\begin{equation}
\psi \rightarrow \psi^{\dagger}\hat{\mathcal{T}};\quad\hat{\mathcal{T}}\equiv i\hat{\tau}_{y}^{AB}\hat{\tau}_{y}^{KK'}\hat{\sigma}_{y}.
\label{eq:TimeReversalSymmetry}
\end{equation}

\subsection{Single particle spectrum}
We write the Hamiltonian for this model as
\begin{equation}
H\equiv H_{0}+H_{C}+H_{int}.
\label{eq:HDef1}\end{equation}
The single-particle part of the Hamiltonian in the two band model\cite{McCann} reads (we will put $\hbar=1$ in all the subsequent formulas)
\begin{equation}
H_{0}\equiv\sum_{k}\psi^{\dagger}\left[\frac{1}{2m}\tau_{z}^{KK'}(\tau_{+}^{AB}k_{+}^{2}+\tau_{-}^{AB}k_{-}^{2})\right]\psi.\label{eq:H0Def}\end{equation}

Here we ignore the "warping term" \cite{McCann} caused by the small skew hopping ($\gamma_3$) since it would have a negligible effect on the RG. 
We have defined $k_{\pm}=k_{x}\pm ik_{y}$, and $m = \frac{2\gamma_1}{gr^2_{AB}\gamma_0^2}$ where $\gamma_0$ is the the interlayer integral and $r_{AB}$ is the interatomic distance.  
(Effect of the electron-electron interaction on the warping was studied in Ref.~\onlinecite{Letter}.)
The Hamiltonian in Eq. (\ref{eq:H0Def}) has the eigenvalue spectrum

\begin{equation}
\varepsilon(\boldsymbol{k}) = \pm \frac{k^2}{2m},
\label{eq:Dispersion}
\end{equation}
where each branch is four-fold (spin and valley) degenerate. The system described by Hamiltonian (\ref{eq:H0Def}) has a higher symmetry than the underlying lattice. This larger symmetry is described by the  $SU(4)\otimes U(1)$ group whose sixteen generators $M_{ij}$ are given by 
\begin{equation}
M_{ij} = \sigma_{i}\tilde{\tau}^{KK'}_{j}\quad(i,j=0,x,y,z),
\label{eq:SU(4)gen}
\end{equation}
\begin{equation}
\begin{aligned}
\tilde{\tau}^{KK'}_0 =\tau^{KK'}_0;&\quad\tilde{\tau}^{KK'}_z =\tau^{KK'}_z,\\
\tilde{\tau}^{KK'}_x =\tau^{AB}_{z}\tau^{KK'}_x;&\quad\tilde{\tau}^{KK'}_y =\tau^{AB}_{z}\tau^{KK'}_y.
\end{aligned}
\end{equation}
An additional rotational $U(1)$ symmetry extends the discrete rotation $\hat{C}_{3}$ 
to a continuous transformation given by $\psi(\boldsymbol{r}) \rightarrow \exp(-2i\theta\sigma_{z}^{AB})\psi(\hat{R}(\theta)\boldsymbol{r})$, where $\hat{R}(\theta)$ is the real space rotation by an angle $\theta$

The preceding discussion actually undercounts the symmetry algebra of
the single particle Hamiltonian greatly, since they do not include the
continuous particle hole symmetry rotations \cite{SymNote}. Including
these rotations, the total symmetry group is $Sp(8)$. However these
extra rotations are not necessary for the following analysis.

\subsection{Electron-electron interactions}

The Coulomb interaction
\begin{equation}
H_{C}\equiv\frac{e^2}{2}
\int\frac{d^{2}{\bf r}d^{2}{\bf r'}}{|{\bf r}-{\bf r'}|}\left[\left(\psi^{\dagger}\psi\right)_{\bf r}\left(\psi^{\dagger}\psi\right)_{\bf r'}\right],\label{eq:Hcoul}\end{equation}
 is the largest interaction energy in the system. The strength of Coulomb interaction on the length scale $L$ is $e^2/L$. The electron kinetic energy related to the same energy scale is $1/(mL^2)$ so the Coulomb interaction will dominate at the scale $L = 1/(me^2)$, which is comparable to
Bohr radius. However due to the generation of electron-hole pairs, the Coulomb interaction is screened, leading to the reduction of interaction energy $e^2/L \rightarrow 1/(mNL^2)$. This screened interactions respects all the symmetries of the system and does not scale; therefore by itself it does not induce any spontaneous symmetry breaking of the lattice symmetry group. 
We will return to the quantitative description of the screened Coulomb interaction in Sec \ref{SEC3}.

Any lattice symmetry breaking is captured by the scaling of the marginal short range interactions. These interactions also reduce the symmetries of the low energy model almost down to the crystal group\cite{AKT,Son2,Letter,myfootnote1}, 
\begin{equation}
H_{int}\equiv\frac{2\pi}{m}\int d^{2}{\bf r}\sum_{ij}g_{ij}\left[\psi^{\dagger}\hat{\tau}_{i}^{AB}\hat{\tau}_{j}^{KK'}\psi\right]_{\bf r}^{2},
\label{eq:DefHSR}
\end{equation}
where have included a factor of $\frac{2\pi}{m}$ to make the couplings dimensionless. The ${\mathcal D}''_{3d}$ symmetry of the two band BLG model forces various relations among the $g_{ij}$:
\begin{equation}
\begin{aligned} 
&g_{xx}=g_{xy}  =g_{yx}=g_{yy}\equiv g_{G}\\
&g_{xz}=g_{yz}\equiv g_{E_{2}};  \,\,g_{zx}=g_{zy}\equiv g_{E_{2}''}\\
&g_{x0}=g_{y0}\equiv g_{E_{1}}; \,\, g_{0x}=g_{0y}\equiv g_{E_{1}''}\\
&g_{z0}\equiv g_{B_{1}};\,\, g_{0z} \equiv g_{A_{2}};\,\, g_{zz}\equiv g_{B_{2}}
\end{aligned}
\label{eq:gIrreps}\end{equation}
Here we have labeled the couplings by the appropriate representation
of $D_{3d}^{''}$ schematically represented in  Fig \ref{fig:Irreps}.
We note for future reference that the interaction terms $g_{E_{2}}(\psi^{\dagger}\tau_{x,y}^{AB}\tau_{z}^{KK'}\psi)^{2}$
and $g_{B_{1}}(\psi^{\dagger}\tau_{z}^{AB}\psi)^2$ are invariant under
the entire $U(4)$ (and which can be extended to $Sp(8)$ by including the particle-hole rotations\cite{SymNote}) symmetry of $H_{0}$.
All other short range interactions, such as those of the form  $\sim(\psi^{\dagger}_\mu\vec{\sigma}_{\mu\nu}\psi_\nu)^{2}$
or $\sim|\psi^{\dagger}_\mu\psi^{\dagger}_\nu|^{2}$
can be always rearranged into the form of $H_{int}$ by using standard
Pauli matrix identity $2\delta_{\mu\nu} \delta_{\mu'\nu'} = \delta_{\mu\mu'}\delta_{\nu\nu'} +\vec{\sigma}_{\mu\mu'} \vec{\sigma}_{\nu'\nu}$.

\begin{figure}
\includegraphics{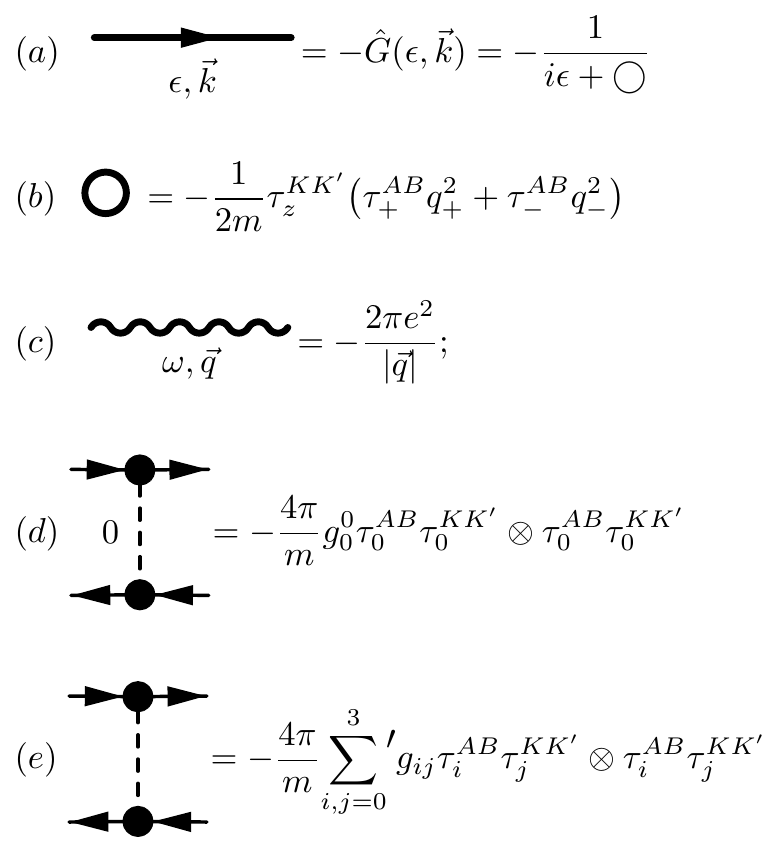}
\caption{Definition of the elements of the diagrammatic expansion. The thick line is the fermion propagator, the circle is the self energy from the single particle of the spectrum. The wavy line is the Coulomb propagator and the dotted line is the contact interaction. We separate the scalar contact interaction $g_{00}$ from the other interactions. }
\label{fig:FeynDefs}
\end{figure}

\section{Perturbation theory and RG equation}
\label{SEC3}
\subsection{$1/N$ resummation}

\begin{figure}
\includegraphics[scale =.92]{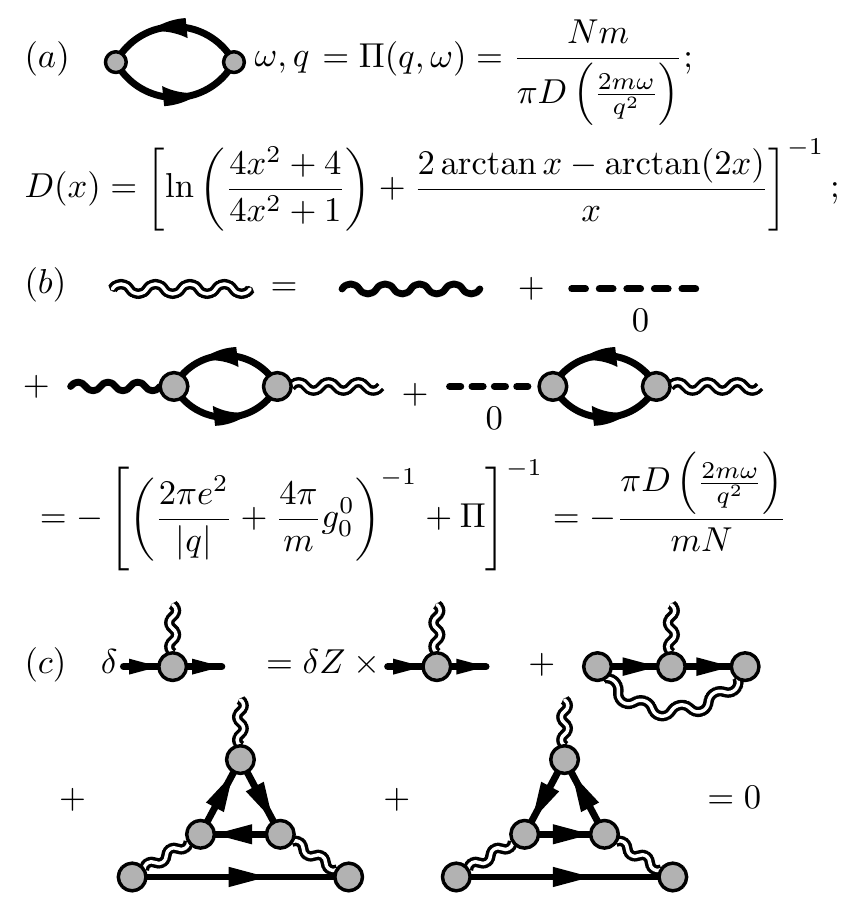}
\caption{Resummation of the strong Coulomb interaction in the $1/N$ approximation. a) Evaluation of the polarization loop. b) Definition of the resummed propagator, represented by the double wavy line. The scalar contact interaction is included in the resummation as it has the same matrix structure as the Coulomb interaction. c) The non renormalization of the Coulomb vertex as a result of gauge invariance. ($\delta Z$ is defined in Fig.~\ref{fig:FeynRG} a).}
\label{fig:FeynRPA}
\end{figure}

For the Coulomb interaction we will use $1/N$ as a small parameter, where $N=4$
is the number of degenerate fermion flavors\cite{Son,AKT,Son2,Letter}. We achieve this expansion by performing
the usual RPA resummation of diagrams (Fig \ref{fig:FeynRPA}). Note that the
coupling $g_{00}$ has the same matrix structure as the long range Coulomb
interaction. We therefore resum the two together,{\it i.e.} we take the
bare interaction in the RPA resummation to be 
\begin{equation}
\mathcal{V}^{(0)}(q)\equiv\frac{2\pi e^{2}}{|q|}+\frac{4\pi g_{00}}{m}.
\label{eq:bareProp}
\end{equation}
Summing up the geometric series of terms in Fig. \ref{fig:FeynRPA}(b) we arrive at
the resummed propagator,
\begin{equation}
\mathcal{D}(q,\omega)=\frac{\mathcal{V}^{(0)}(q)}{1+\mathcal{V}^{(0)}(q)\Pi(q,\omega)},\label{eq:Propogator}\end{equation}
 where
 \begin{equation}
\begin{aligned}\Pi(q,\omega) & =\frac{mN}{\pi f(\frac{2m\omega}{q^{2}})},\\
f(x)\equiv & \left[\log\left(\frac{x^{2}+1}{x^{2}+1/4}\right)+\frac{2\arctan x-\arctan2x}{x}\right]^{-1}.\end{aligned}
\label{eq:Polarization}\end{equation}

We further take the long wavelength limit, $q\rightarrow0$, where $V^{(0)}(q)\Pi\gg1$.
This gives us the approximate expression for the interaction propagator,
\begin{equation}
\mathcal{D}(q,\omega)\approx \frac{1}{\Pi(q,\omega)}= \frac{\pi}{mN}f\left(\frac{2m\omega}{q^{2}}\right).
\label{eq:ReducedProp}
\end{equation}
Since $\mathcal{D}\propto1/N$ we can use a perturbative expansion in $1/N$. Note that we have neglected the higher-energy bands in considering the resummation of the Coulomb potential.
However, the higher energy bands would only change the dielectric constant which cancels out of the final formula.

Now we write the partition function as a path integral in imaginary time $t$ over Grassman fields $\psi$ and $\psi^{\dagger}$,
\begin{equation}
\begin{aligned}
Z = \int& D\psi D \psi^{\dagger} e^{-S} \\
S\equiv \int d^{2}rdt& \left(\psi^{\dagger}\frac{d}{dt}\psi - H\left[\psi^{\dagger},\psi\right]\right),
\end{aligned}
\label{eq:Action}\end{equation}
where $H$ is defined in equation (\ref{eq:HDef1}). Then, we perform the RG by integrating out all fermionic states with momenta $q$ such that
$K>|q|>K\: e^{-\ell}$ , where $K$ is some ultraviolet cutoff regardless
of $\omega$. We will set $K_{0}$ so that $K_{0}^2 / (2m) =\gamma_1/2$, approximately
the upper-limit of the applicability of the two band model with the parabolic dispersion.
We then rescale $\psi\rightarrow(1+\delta Z/2)\psi$ to keep the term
$\int\psi^{\dagger}\frac{d}{dt}\psi$ unchanged. This procedure has the
benefit of not renormalizing the Coulomb vertex because of gauge
invariance (see Fig. \ref{fig:FeynRPA}(c)). If we assign $t$ an RG dimension $2$
then at tree level the operator $\psi$ has RG dimension $+1$, and $m$,
$g_{ij}$ and the Coulomb interaction are marginal.  

\begin{figure}
\includegraphics{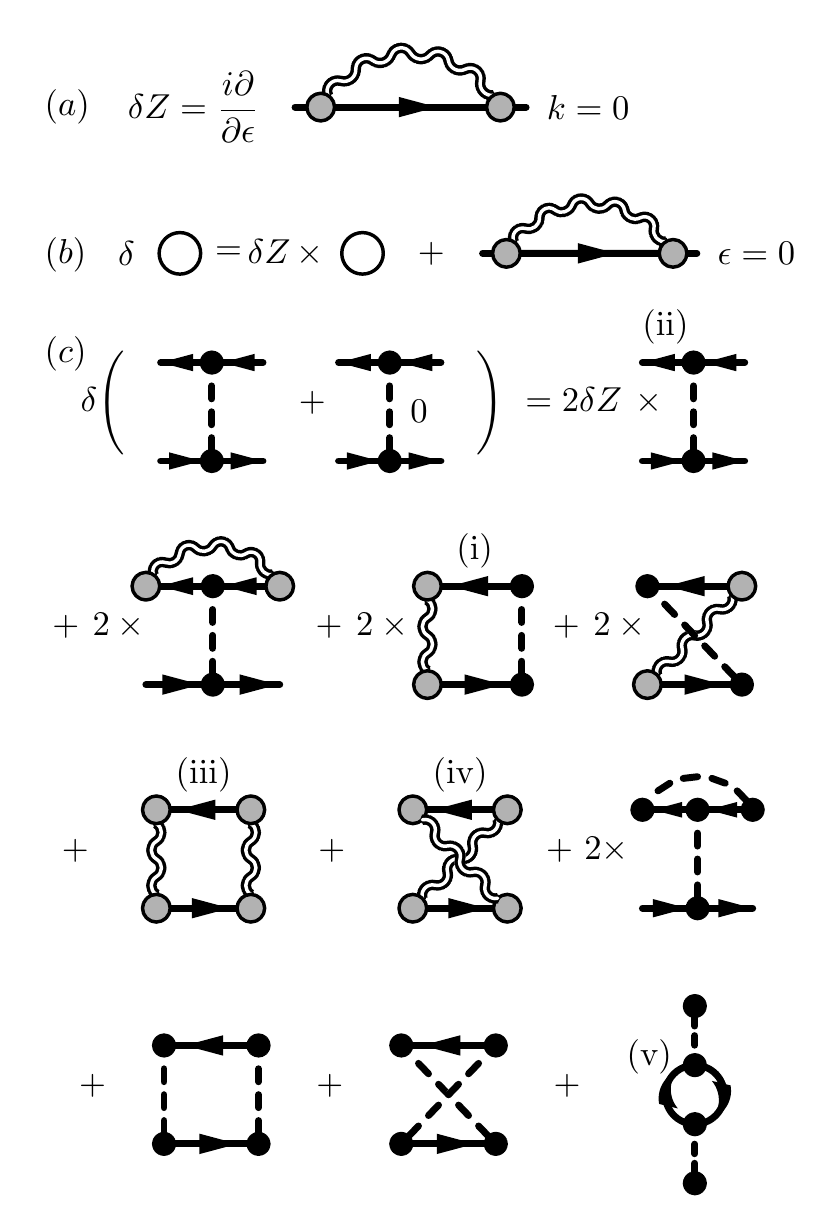}
\caption{Diagrams included in RG equations. a) Single-particle particle weight renormalization; b) Renormalization of the mass $m$. c) Renormalization of the contact interactions.}
\label{fig:FeynRG}
\end{figure}

There is a subtlety in the $1/N$ treatment of the Coulomb interaction. Because of the behavior of the interaction in the limit $q\rightarrow0$, $\omega\rightarrow\infty$, some of the diagrams taken individually diverge faster than logarithmically.
For example, the self energy diagram (see Fig. \ref{fig:FeynRG}(a))
gives the correction to the quasiparticle weight,
\begin{eqnarray}
\delta Z & = & \frac{\partial}{i\partial\Omega}|_{\Omega=0}\int\frac{d\omega d^{2}\boldsymbol{k}}{(2\pi)^{3}}\left(\frac{-\pi}{mN}\right)f\left(\frac{2m\omega}{k^{2}}\right)\nonumber \\
 &  & \times\left[\frac{-i(\omega+\Omega)+\hat{\tau}^{KK'}_{z}(\hat{\tau}^{AB}_{+}k_{+}^{2}+\hat{\tau}^{AB}_{-}k_{-}^{2})}{(\omega+\Omega)^{2}+(\frac{k^{2}}{2m})^{2}}\right]\nonumber \\
  & = & \frac{1}{N}\log K\,\int_{-\infty}^{\infty}\frac{dx}{2\pi}f(x)\frac{1-x^{2}}{(1+x^{2})^{2}},\label{eq:FieldStrength}\end{eqnarray}
where the variable $x$ is defined by the substitution $\omega=xk^{2}/(2m)$. This integral is formally
infinite since $f(x)\rightarrow x$ as $x\rightarrow\infty$. To understand this divergence, note that it comes from the region where
the momentum $k$ through the Coulomb line goes to zero. This corresponds to a spatially constant but time-varying potential $V(t)$. Such
a potential is merely a constant shift in energy so that the Green's function are changed as $G\left(t_1,t_2\right)\rightarrow G\left(t_1,t_2\right)\exp\left(ie\int_{t_1}^{t_2}V\left(t\right)dt\right)$. It is the summation over the fluctuations of this phase that produces the divergence.
 However, in all observables gauge invariant quantities the fermion lines must come in closed loops which cancels out this phase, and so it cannot appear in any physical quantities.  Reassuringly in all our calculations this is the case. Indeed such a  divergence cancels out from the correction to the electron mass,
\begin{equation}
\delta\left(\frac{1}{2m}\right) = \frac{1}{2}\frac{\partial^2}{\partial p_{+}^2}\Sigma- \frac{1}{2m}\delta Z,
\label{eq:RGisFinite}
\end{equation}
where
\begin{equation}
\begin{aligned}
&\frac{\partial^2}{\partial p_{+}^2}\Sigma=\\
 & \, tr\Bigg\{\frac{1}{2}\hat{\tau}^{KK'}_{z}\hat{\tau}^{AB}_{-}\frac{\partial}{\partial p_{+}^{2}}|_{p=0}
\int\frac{d\omega d^{2}\boldsymbol{k}}{(2\pi)^{3}}\frac{\pi}{mN}\left[-f\left(\frac{2m\omega}{k^{2}}\right)\right]\\
 & \times\hat{G}(\omega,\boldsymbol{k}+\boldsymbol{p)}\Bigg\}\\
= & \,- tr\Bigg\{\frac{1}{2}\hat{\tau}^{KK'}_{z}\hat{\tau}^{AB}_{-}\frac{\partial}{\partial p_{+}^{2}}|_{p=0}\int\frac{d\omega d^{2}\boldsymbol{k}}{(2\pi)^{3}}\frac{\pi}{mN}f\left(\frac{2m\omega}{k^{2}}\right)\\
 &\!\!\times \frac{-i\omega+\frac{1}{2m}\hat{\tau}^{KK'}_{z}(\hat{\tau}^{AB}_{+}(k_{+}+p_{+})^{2}+\hat{\tau}^{AB}_{-}(k_{-}+p_{-})^{2})}{\omega^{2}+
\left[(
\frac{\left(\boldsymbol{k}+\boldsymbol{p}\right)^2}
{2m}\right]^{2}}\Bigg\}\\
= & \,-\!\!\!\int\!\!\!\frac{d^{2}k}{(2\pi)^{2}}\frac{1}{k^{4}}\!\int\!\!\frac{dx}{2\pi}\frac{\pi f(x)}{mN}\frac{\partial}{\partial p_{+}^{2}}|_{p=0}\frac{k_{+}^{2}+2k_{+}p_{+}+p_{+}^{2}}{x^{2}\!+\!(1+\frac{k\cdot p}{k^{2}}+\frac{p^{2}}{k^{2}})^{2}}\\
 = & \,-\frac{1}{2mN}\log K\,\int\frac{dx}{(2\pi)}f(x)\frac{x^{4}-3x^{2}}{(x^{2}+1)^{3}}.\end{aligned}
\label{eq:MassDiagram}\end{equation}

Although the latter expression is divergent in the limit $x\rightarrow\infty$, the sum,
\begin{equation}
\delta\left(\frac{1}{2m}\right)  =\frac{1}{2mN}\log\,K\int_{-\infty}^{\infty}\frac{dx}{2\pi}f(x)\frac{1-3x^2}{\left(1+x^2\right)^3},
\label{eq:RGisFinite2}
\end{equation}
is convergent. Therefore the mass has a logarithmic dependence on cutoff, as expected. This enables us to write down the RG equations for the electron mass  $m$, 
\begin{equation}
\frac{d\log m(\ell)}{d\ell}=-\frac{\alpha_{1}}{2N};\quad\ell\equiv\log(K_0/K),\label{eq:MassRG}\end{equation}
where  
\begin{equation}
\alpha_{1}\equiv\frac{1}{2\pi}\int dx\, f(x)(1-3x^{2})/(1+x^{2})^{3} \approx -0.078.
\label{eq:defAlpha1}
\end{equation}
Since $\alpha_{1}/(2N) <10^{-2}$ is very small we shall neglect this mass renormalization for the rest of this analysis. 

\subsection{Renormalization of the contact interactions}

We now consider the renormalization of the short-range interactions. 
Based on our assumption that the bare values $g_\mathcal{A}$ are small we will work to order $g^{2}$
and to lowest order in $1/N$. 

The leading logarithmic corrections to the coupling constants of the contact interaction (\ref{eq:DefHSR}) are shown on the Fig.~\ref{fig:FeynRG}(c). Straightforward calculation of those diagrams yield\cite{myfootnote2}
the set of RG equations for the 8 coupling constants $g_\mathcal{A}$
\begin{equation}
\begin{aligned}\frac{dg_{ij}}{d\ell}= & \!-\frac{\alpha_{3}}{N^{2}}\delta(E_{2})_{ij}-\frac{\alpha_{1}\!+\!2\alpha_{2}A_{ij}}{N}g_{ij}\\
-\sum_{kl}^{\sim} & \frac{g_{kl}}{N}\alpha_{2}B_{ij}^{kl}-2NA_{ij}g_{ij}^{2}+\sum_{kl}^{\sim}\sum_{mn}^{\sim}C_{klmn}^{ij}g_{kl}g_{mn},\end{aligned}
\label{eq:gRG}\end{equation}
where
\begin{equation}
\begin{aligned}
A_{ij}\equiv&-\frac{1}{16}\sum_{\gamma=x,y}tr\left(\left[\hat{\tau}^{KK'}_{i}\hat{\tau}^{AB}_{j},\hat{\tau}^{KK'}_{z}\hat{\tau}^{AB}_{\gamma}\right]^{2}\right),\\
B_{kl}^{ij}\equiv&\frac{1}{64}\sum_{\gamma=x,y}tr\left(\hat{\tau}^{KK'}_{k}\hat{\tau}^{AB}_{l}\left\{ \hat{\tau}^{KK'}_{i}\hat{\tau}^{AB}_{j},\hat{\tau}^{KK'}_{z}\hat{\tau}^{AB}_{\gamma}\right\} \right)^{2},\\
C_{klmn}^{ij} &=\\
\frac{1}{8} & \!\sum_{\gamma=x,y}tr\Big(\hat{\tau}^{KK'}_{k}\hat{\tau}^{AB}_{l}\hat{\tau}^{KK'}_{i}\hat{\tau}^{AB}_{j}\hat{\tau}^{KK'}_{z}\hat{\tau}^{AB}_{\gamma}\\
&\qquad\quad\, \times\left[\hat{\tau}^{KK'}_{k}\hat{\tau}^{AB}_{l},\hat{\tau}^{KK'}_{z}\hat{\tau}^{AB}_{\gamma}\right]\hat{\tau}^{KK'}_{i}\hat{\tau}^{AB}_{j}\Big)\\
+\frac{1}{64}&\!\sum_{\gamma=x,y}\Big\{tr\Big(\hat{\tau}^{KK'}_{i}\hat{\tau}^{AB}_{j}\Big[\hat{\tau}^{KK'}_{k}\hat{\tau}^{AB}_{l}\hat{\tau}^{KK'}_{z}\hat{\tau}^{AB}_{\gamma}\hat{\tau}^{KK'}_{m}\hat{\tau}^{AB}_{n}\\
&\qquad\quad\,+ \hat{\tau}^{KK'}_{k}\hat{\tau}^{AB}_{l}\hat{\tau}^{KK'}_{z}\hat{\tau}^{AB}_{\gamma}\hat{\tau}^{KK'}_{m}\hat{\tau}^{AB}_{n}\Big]\Big)\Big\}^2\\
+\frac{1}{32}&\left\{tr\left(\hat{\tau}^{KK'}_{i}\hat{\tau}^{AB}_{j}\left[\hat{\tau}^{KK'}_{k}\hat{\tau}^{AB}_{l},\hat{\tau}^{KK'}_{m}\hat{\tau}^{AB}_{n}\right]\right)\right\}^2.
\end{aligned}
\label{eq:DefTensors}\end{equation}
Here $\tilde{\sum}_{ij}$ is a sum over $i,j=\{0,x,y,z\}$ excluding the
combination $i=0,j=0$ and the summation convention is not used. The symbol $\delta(E_{2})_{ij}$
is 1 when $i=z$ and $j=x,y$ and 0 otherwise. By appearance there are 16 equations contained in Eq. (\ref{eq:gRG}). However several of these are identical due to the $\mathcal{D}_{3d}''$ symmetry so there are only eight independent equations for the flow of the eight independent coupling constants. 
The numerical coefficient $\alpha_1$ is defined in Eq. (\ref{eq:defAlpha1}) and
\begin{eqnarray}
\alpha_{2}\equiv\int\frac{dx}{2\pi}\frac{2f(x)}{(1+x^{2})^{2}}\approx.469,\\
\alpha_{3}\equiv\int\frac{dx}{2\pi}\frac{f(x)^{2}}{4(1+x^{2})^{2}}\approx.066. \label{eq:defAlpha3}
\end{eqnarray}

The term $2NA_{ij}g^2_{ij}$ in Eq. (\ref{eq:gRG}) corresponding to leading loop diagram (v) in Fig. \ref{fig:FeynRG}(c) is naively the most significant quadratic term in Eq. (\ref{eq:gRG}), because it is leading in $N$. This term represents screening of repulsive interactions in the charge channel as expected in a fermionic system (since $A_{ij} \ge 0$). Note that this term is actually zero for the representation $E_2''$ and $A_2$, because these interactions commute with the single particle Hamiltonian. Therefore, to lowest order in $1/N$, the interactions $E_2''$ and $A_2$ are unscreened and free to grow strongly attractive. We hasten to add that that the higher order terms in Eq. (\ref{eq:gRG}) are very important and one cannot understand the behavior of the RG flows based only on the leading terms. 

The single particle Hamiltonian is off-diagonal so there is a contribution of order $g^{0}N^{-2}$ to the coupling $g_{E_{2}}$
from the two Coulomb line diagram in Fig. \ref{fig:FeynRG}(c)(iii,iv). These may be calculated, 
\begin{equation}
\begin{aligned} \int&\frac{d^{2}kd\omega}{(2\pi)^{3}}\left[\frac{\pi f(\frac{2m\omega}{k^{2}})}{Nm}\right]^{2}\hat{G}(k,\omega)\otimes\left(\hat{G}(k,\omega)+\hat{G}(k,-\omega)\right)\\
= & \,2\!\int\frac{d^{2}kd\omega}{(2\pi)^{3}}(\frac{\pi f(\frac{2m\omega}{k^{2}})}{Nm})^{2}(\frac{k^{2}}{2m})^{2}\\
&\quad\times\,\left(\frac{\tau^{KK'}_{z}\tau^{AB}_{+}\otimes\tau^{KK'}_{z}\tau{AB}_{-}+hc}{(\omega^{2}+(\frac{k^{2}}{2m})^{2})^{2}}\right)\\
= & \frac{\pi}{N^{2}m}\int\!\!\frac{dk}{k}\!\!\int\!\!\frac{dx}{2\pi}\frac{f(x)^{2}}{(1+x^{2})}\!\sum_{\gamma=x,y}\!\left(\tau^{KK'}_{z}\tau^{AB}_{\gamma}\otimes\tau^{KK'}_{z}\tau^{AB}_{\gamma}\right)\\
= & \frac{4\pi\alpha_{3}}{N^{2}m}\log K\:\sum_{\gamma=x,y}\left(\tau^{KK'}_{z}\tau^{AB}_{\gamma}\otimes\tau^{KK'}_{z}\tau^{AB}_{\gamma}\right).\end{aligned}
\label{eq:CrossedCoulomb}\end{equation}

From this, it follows the that the free field point $g_{\mathcal{A}}=0$ is not a fixed point. Even if the the system starts with all bare
couplings $g_\mathcal{A}(\gamma_1/2)=0$ it will flow under RG to have finite $g_{E_{2}}$
and $g_{B_{1}}$ with the other couplings fixed to zero by the
$SU(4)$ symmetry of the single particle Hamiltonian. To demonstrate
the behavior in this regime we ignore momentarily $g_{B_{1}}$ which
gives us a single equation for $g_{E_2}$,
\begin{equation}
\begin{aligned}\frac{dg_{E_{2}}(\ell)}{d\ell}=-\frac{1}{N(N+2)} & \left(\frac{\alpha_{3}(N+2)}{N}-\frac{(\alpha_{2}-\alpha_{1})^{2}}{8N}\right)\\
-\,2(N+2) &\left (g_{E_{2}}-\frac{\alpha_{2}-\alpha_{1}}{4N(N+2)}\right)^{2}.\end{aligned}
\label{eq:SimpleRG}\end{equation}
Since the first term on the RHS is negative, there can be no fixed
point and $g_{E_{2}}$ flows to $-\infty$ regardless of the initial
conditions. (This holds whether we treat Eq. (\ref{eq:SimpleRG}) to lowest order in $N$ or simply plug in $N=4$). According to the mean-field theory (see Sec. \ref{SEC4}),
this suggests a nematic ground state\cite{myfootnote2} with transition at $\approx 100$mK.

\subsection{Applicability of our approximations}

\begin{figure}
\includegraphics[width=.7\columnwidth]{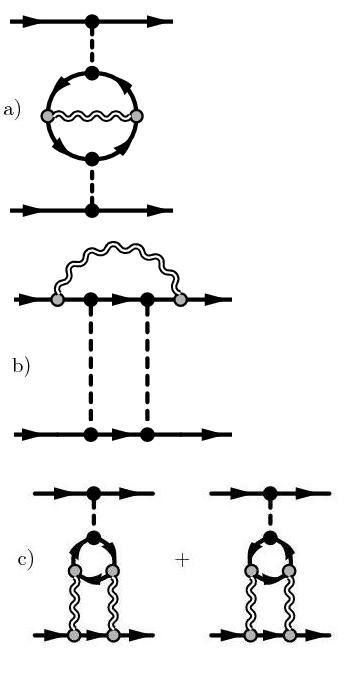}
\caption{a) Schematic representation of the "bubble" diagrams where the shaded blob represents all possible connected diagrams, and arbitrary Coulomb propagators may be added. b) Similar representation of the "ladder" diagrams.
 The leading diagrams from both of these groups are included, even though this is not strictly parametrically correct. c) A second loop contribution to the anomalous dimension of the coupling $g_{E_2}$ which is disregarded.}
\label{fig:FeynDiag2}
\end{figure}

Let us turn to the justification of only including the diagrams Fig \ref{fig:FeynRG}(c) in our treatment. Notice that it is different from the conventional $1/N$ approximation, see e.g Ref.~\onlinecite{Coleman}. There are two issues:
(1) there are two loop diagrams which are leading order in $1/N$ but are not included, Fig. \ref{fig:FeynDiag2}(c); and
(2) there are diagrams that are subleading in $N$ which are taken into account (compare the bubble and ladder diagrams in Fig. 
\ref{fig:FeynRG}(c).

To address the first issue, let us discuss diagrams of Fig. \ref{fig:FeynDiag2} c in more detail. They are non-vanishing only for $g_{E_2}$  and have the form $\delta g_{E_2}\sim \frac{g_{E_2}}{N}\left(C\ln^2(K) +D\ln(k)\right)$. The term $\ln^2$ is produced by two iterations of the the RG equations (by substituting diagrams (iii) and (iv) into diagram (v)) but the second term does contribute to the linear term in the RG equation $2\alpha_2 A_{E_2}\rightarrow 2\alpha_2 A_{E_2}+D$.  
The constant $D$, however, depends on the cutoff scheme so that the term linear in $g_{E_2}$ in the RG equation for $g_{E_2}$ is not known (for the other constants it is well defined). Fortunately, is does not matter for the
divergent behavior at large $N$. Consider the situation with all other constants except $g_{E_2}$ fixed to zero, keeping only coefficients leading in $1/N$, compare Eq. (\ref{eq:SimpleRG}),
\begin{equation}
\frac{dg_{E_2}}{d\ell} = -\frac{\alpha_3}{N^2} -\frac{\alpha_1 +2\alpha_2}{N}g_{E_2} -2Ng_{E_2}^2\label{eq:leadgE2RG}\end{equation}
 The quadratic term dominates the constant term when $g_{E_2}\ge N^{-\frac{3}{2}}$ at this point, but then the linear term 
is smaller by a factor of $1/\sqrt{N} \ll 1$. Thus, contrary to initial appearance, the linear term is of higher order in $1/N$  for $g_{E_2}$ - 
so that we leave it in Eq. (\ref{eq:gRG}) only for simplicity. It makes essentially no difference to the evolution of the RG equations. 

To address the second question we notice that the bubble diagram Fig. \ref{fig:FeynRG}c(v) contains an extra factor of $N$ in 
comparison with diagrams (vi, vii,viii). The latter diagrams are not diagonal in terms of the coupling constant, as given by the tensor 
$C^{ij}_{klmn}$, whereas the bubble diagram is $\propto N g_{ij}^2$ by construction. The large amounts of constants involved in the 
non-diagonal term may overcome the factor of $N$ in the diagonal terms; therefore keeping both is legitimate. The higher order terms 
may be considered as $1/N$ corrections to the tensors $A_{ij}$ and $C^{ij}_{klmn}$ respectively. For example, Fig. \ref{fig:FeynDiag2}(a) is a
 leading $1/N$ correction to $A_ij$, whereas Fig. \ref{fig:FeynDiag2}(b) is a leading $1/N$ correction to $C^{ij}_{klmn}$, even though the two diagrams do not have the same in order in $N$.

Finally, we compare our treatment to the existing theoretical contributions. The work of Vafek and Yang \cite{Vafek} is similar in spirit
 but contains only the $G_1$ and  $B_2$ out of the eight possible  representations and treats the Coulomb interaction as short 
range. The later work of Vafek\cite{Vafek2}contains the RG equations for the full eight constants but again treats the Coulomb interaction as short ranged, and does not attempt to describe the general structure of the RG flow. 
The treatment of Ref.~\onlinecite{Levitov} is completely at the mean-field level 
and corresponds to counting only the diagrams from Fig.~\ref{fig:FeynRG}c marked (i) and (ii), which is not a parametrically 
justified approximation as well as considering only the $B_2$ representation. The RG equation of Ref.~\onlinecite{MacDonald2} 
essentially ignore the spin-valley structure but still does not retain the necessary number of coupling constants to describe even that limited situation.

\section{RG Flows, Their Termination and Renormalized Mean Field Treatment of Symmetry Breaking}
\label{SEC4}
In this section we describe the numerical analysis of the coupling constant RG flows described by Eq. (\ref{eq:gRG}) and show that there are no weak coupling fixed points. The divergence of coupling constants in 2D at zero temperature indicates spontaneous symmetry breaking. (Unlike in 1D the quantum fluctuations in 2D are not infrared divergent and do not destroy zero temperature phases.) We analyze the resulting phases within mean-field theory, using the coupling constants renormalized by the RG. This treatment is superior to simply doing mean field start from the high energy scale since in that case the large logarithms are not summed in a controlled fashion.

\subsection{General structure}
\begin{figure}
\includegraphics[scale  = .6]{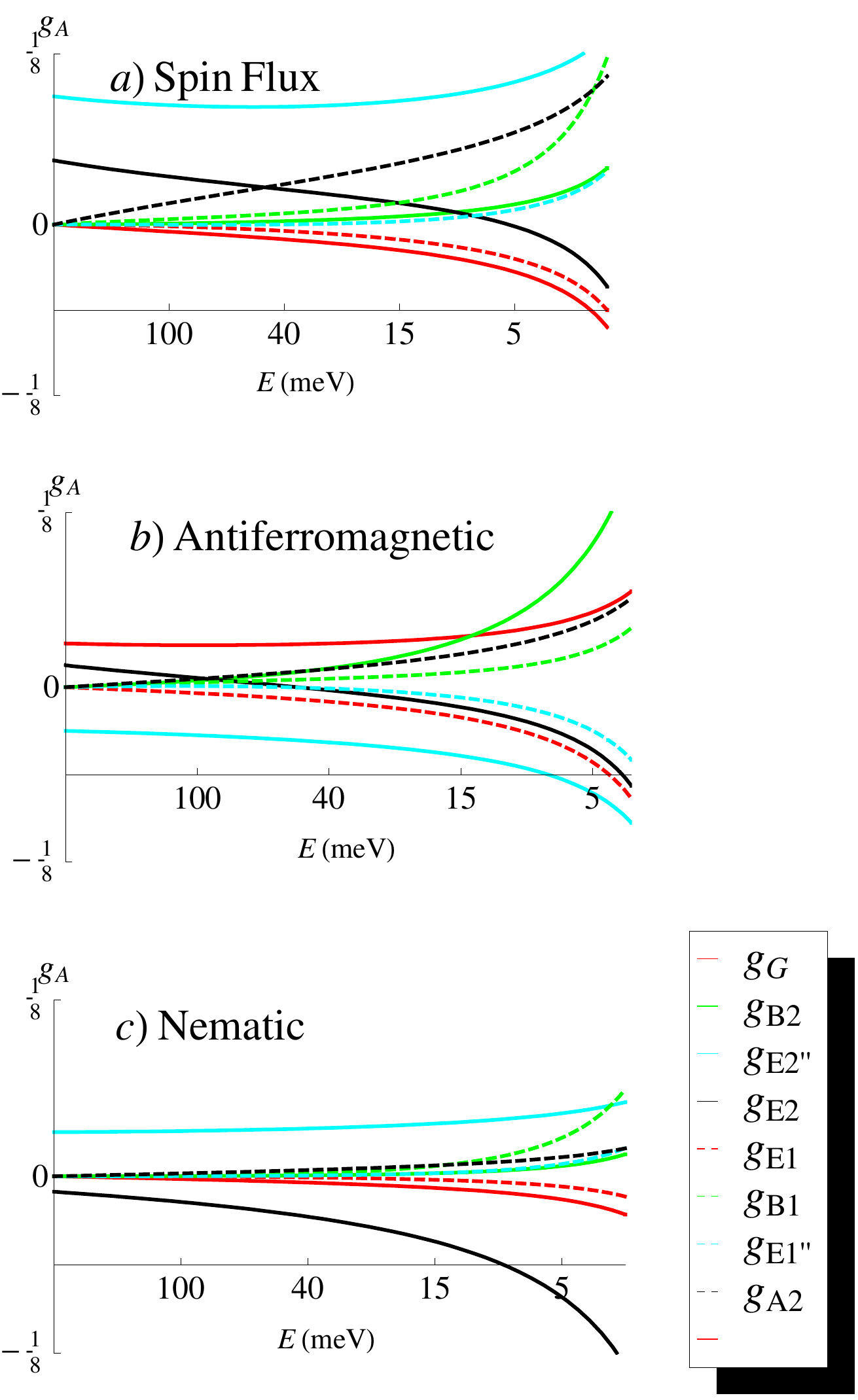}
\caption{ (Color online)Plots of the coupling constant as a function the running RG scale $\ell = \log(K_0/K)$. There are eight running couplings labeled by the corresponding representation. The density-density couplings are given by solid lines, the current-current couplings by dashed. The graphs end when the couplings become of order $1/N = 1/4$. They reach a singularity a finite $\ell$ soon after the graph ends. }\label{fig:RGRunning}\end{figure}

If the initial RG conditions are such that $g_{\mathcal{A}}(\gamma_1/2)\neq0$, then the $SU(4)$ symmetry is absent and
 we must consider the flow of all the coupling under the RG. Determining
if there exist any fixed points cannot be done analytically as it requires
solving a polynomial of the 64th order. However, a numerical solution
shows that there exist no fixed points. Therefore at least some of the couplings
must grow infinitely. At the same time, the leading term quadratic in the couplings
in Eq. (\ref{eq:gRG}) $\sim Ng^{2}$, always with a non-positive coefficient,
so we expect generically that large positive coupling to be driven back
to zero. For all positive initial coupling constants,
this means that the system will be driven to the free field point until $g_{E_2}$ becomes large and negative.
 This behavior
is confirmed by the numerical evolution of the RG equations (see Fig. \ref{fig:RGRunning} where $g_{E_2}$ always becomes negative and increases until the other couplings diverge).

Note that although we have set the current-current couplings $g_{B_1}$, $g_{A_2}$, $g_{E_1}$ and $g_{E_1''}$ to zero initially, they are generated through renormalization. The examples of the RG equation shown in Fig. \ref{fig:RGRunning} indicate the current-current couplings become of the same order as the density-density couplings at low energies. Therefore we cannot ignore the current-current interactions when analyzing the ground state, and ignoring them would lead to misleading results.

\begin{figure*}
\includegraphics[width =1.9\columnwidth]{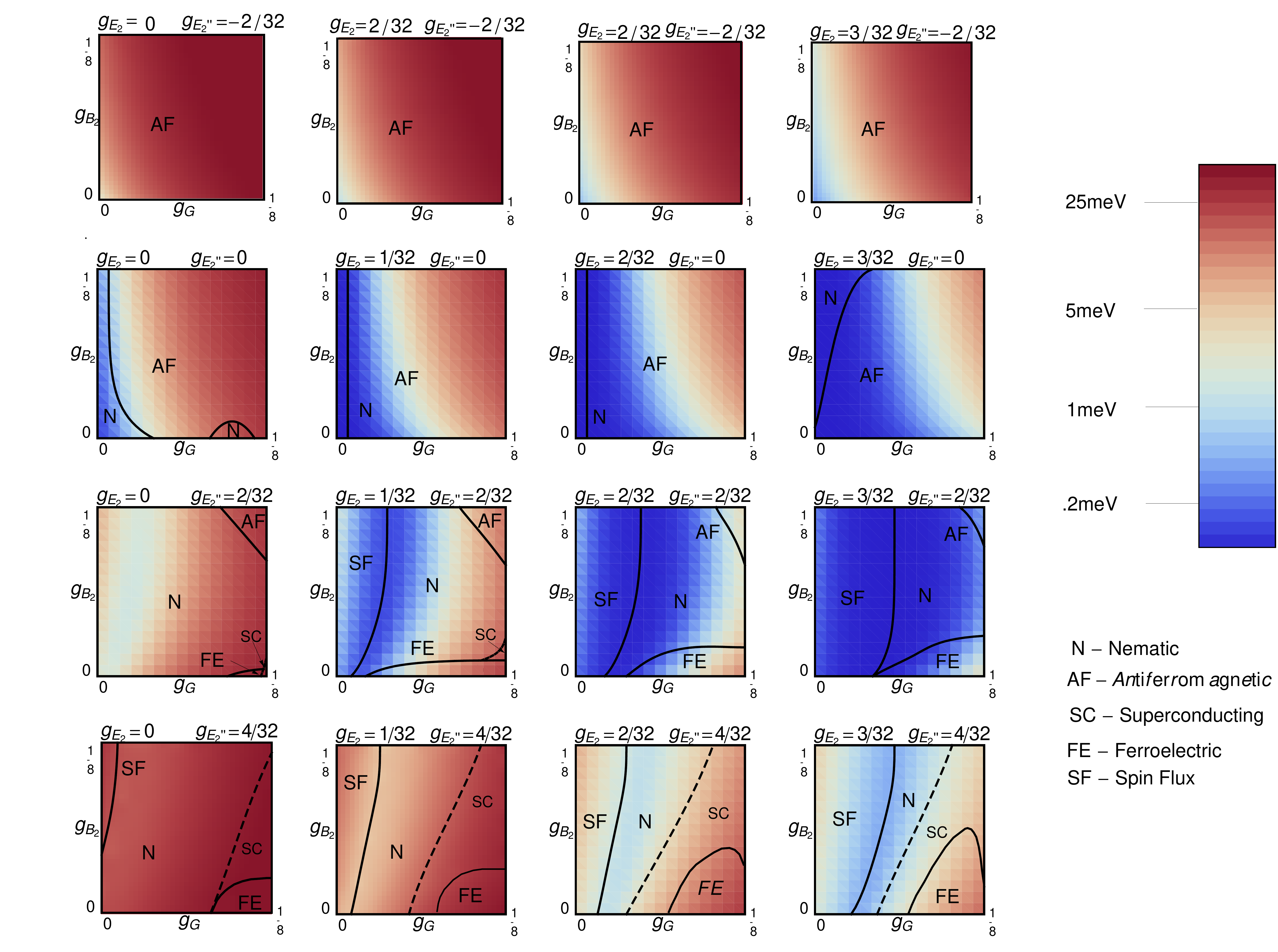} 
\caption{ (Color online)Plot of the broken symmetry phase and energy scale for BLG as a function of bare coupling constants. When the energy scale is less $E_{LiTr} \approx 1 meV$ 
the broken symmetry state is in competition with the Lifshitz transition 
and the BLG may remain metallic (with eight Dirac points). Dashed line indicates the region where
the triplet superconducting phase (SC) is very close in energy (but slightly above)
 to normal (nematic or ferroelectric) states.}
\label{fig:bigPlots}
\end{figure*}

The coupling $g_{B_2}$ has been given special emphasis in some of the earlier studies\cite{Kharitonov}. We find that in the RG equations it does not seem to play an exclusive role, as can be seen in Fig. (\ref{fig:bigPlots}), where it is screened efficiently - the leading term in the RG flow is $2Ng^2_{B_2}$. Note that large positive initial $g_{B_2}$ does not provoke a phase transition to the AF state on its own (see Fig (\ref{fig:RGRunning})).  Once $g_{E_2}$ becomes relatively large the presence of a finite $g_{B_2}$ will change the structure of the flow, especially since it breaks the $SU(4)$, however not in a marked way. For example, starting with all other couplings set to zero except for $g_{B_2}$,  $g_{E_2}$ still becomes the most significant negative coupling, and the nematic phase is the preferred phase. As a result, $g_{B_2}$ is perhaps the least important of the four couplings. This is not a conclusion that can be reached on general grounds, but only by solving the detailed RG equations over a broad range of parameters. Moreover, at least some of the couplings behave non-monotonically.
Initially negligible coefficients may end up diverging quickly (e.g.  $g_{A_2}$ in Fig(\ref{fig:RGRunning}(a)). 
At the same time a couplings that is not large at the end of the RG flow may change the character of the flow in the initial stages.

The RG equations contain terms up to second order in $g$. Therefore, it may be easily seen that since there is no fixed point, the couplings always go to infinity as $g_\mathcal{A} \sim \lambda_\mathcal{A} (\ell_0 - \ell)^{-1}$ where  $\ell_0$ gives the value of the singularity in the RG flow and the $\lambda_\mathcal{A}$ determine how quickly each constant diverges. Mathematically, there are six sets of $\{\lambda_\mathcal{A}\}$ that satisfy the RG equations and are stable to perturbation.  One might be tempted to determine the ground state, using this mathematical feature, via the coefficients $\lambda_\mathcal{A}$. However these coefficients are meaningful only for the RG at $g_\mathcal{A}\gg1$ which is outside the range of validity of the proposed theory and the RG equation (\ref{eq:gRG}). Moreover, $g\rightarrow \infty$, indicates an instability towards a broken symmetry states, so that we shall 
use a mean field theory starting from the energy scale where some of the couplings become sufficiently large.

\subsection{Ground state energies within renormalized mean-field approach}
The unbounded growth of coupling constants
in the RG flow generally indicates the development of a spontaneous
symmetry breaking and the opening of a gap. To describe the corresponding phase transition we use a self-consistent
mean-field theory. The self-consistent mean-field theory is implemented by replacing
all possible pairs of fermions in the quartic interaction terms with their
mean values.  For this we introduce the Gorkov-Nambu vectors which adjoin the two 8-component vectors $\psi$ and $\psi^\dagger$ as follows,
\begin{equation}
\begin{aligned}
\Psi(\boldsymbol{k}) \equiv& \left(\begin{array}{c}
\psi(\boldsymbol{k})\\
\hat{\mathcal{T}}\left(\psi^{\dagger}(-\boldsymbol{k})\right)^{t}\end{array}\right)_N,\\
\Psi^{\dagger}(\boldsymbol{k}) \equiv& \left(\begin{array}{cc}
\psi^{\dagger}(\boldsymbol{k}),&
-\left(\psi(-\boldsymbol{k})\right)^{t}\hat{\mathcal{T}}\end{array}\right)_N,
\end{aligned}\label{eq:NambuDef}
\end{equation}
where $\hat{\mathcal{T}}\equiv i\tau^{KK'}_y\tau^{AB}_y\sigma_y$ is the time reversal matrix. We also introduce Pauli matrices $\tau^{N}_{0,x,y,z}$ acting on the Nambu space. The vectors $\Psi^\dagger$ and $\Psi$ satisfy the condition 
\begin{equation}
\Psi^\dagger(\boldsymbol{k}) = i\Psi^t(-\boldsymbol{k})\tau^N_y\hat{\mathcal{T}}
\label{eq:NambuReal}\end{equation}
We  rewrite the interaction terms with the help of the Nambu vectors
\begin{equation}
 \sum_{ij}\left(\psi^\dagger\tau^{KK"}_i\tau^{AB}_j\psi\right)^2
 =\frac{1}{4}\sum_{s}g_{s}\left(\Psi^{\dagger}\!\cdot\!\hat{M}_s\!\cdot\!\Psi\right)^2.
\label{eq:NambuInt}\end{equation}
Here $M_s \equiv \tau^{KK'}_i\tau^{AB}_j\tau^{N}_k\sigma_l$  acts on the 16 dimensional space spanned by the Nambu vectors and we write $s$ as a shorthand for the list (ijkl). The couplings $g_s$ are defined as $g_{abz0} = g_{ab}$, 
$g_{00z0} = g_{00}$, 
$g_{a000} = g_{a0}$ and $g_{0b00}=g_{0b}$,  where $a,b =x,y,z$ and all other constants are zero.
 The factors of $\tau^N_z$ are necessary since we must have
\begin{equation}
(i\tau^N_y\mathcal{T})M_s^t(i\tau^N_y\mathcal{T}) = - M_s
\label{eq:NambuM}\end{equation}
to satisfy both Eq. (\ref{eq:NambuReal}) and fermion anticommutivity.

The mean field approximation consists of replacing pairs of fermionic  operators in Eq. (\ref{eq:NambuInt}) with their expectation values as follows:
\begin{equation}
\begin{aligned}
\frac{1}{4}\sum_{s}&g_{s}\left(\Psi^{\dagger}\!\cdot\!\hat{M}^{s}\!\cdot\!\Psi\right)^2
\approx\frac{1}{2}\sum_{s}g_{s}\Bigg\{\Psi^{\dagger}\!\cdot\!\hat{M}^{s}\!\cdot\!\Psi\langle\Psi^{\dagger}\!\cdot\!\hat{M}^{s}\!\cdot\!\Psi\rangle\\ 
+2\,&\Psi^\dagger\!\cdot\left(\hat{M}_s\!\cdot\langle\Psi\!\otimes\!\Psi^\dagger\rangle\!\cdot\hat{M}_s\right)\Psi
-\frac{1}{2}\left(\langle\Psi^\dagger\!\cdot\!\hat{M}^{s}\!\cdot\!\Psi^\rangle\right)^2\\
+\,\,&tr\left[\left(\langle\Psi\otimes\Psi^\dagger\rangle\!\cdot\!\hat{M}_s\right)^2\right]\Bigg\}.\\
\end{aligned}
\end{equation}
Here we have used the fact that, according to Eqs.~(\ref{eq:NambuReal}) and (\ref{eq:NambuM}),
\[
\Psi^{\dagger}\!\cdot\!\hat{M}^{s}\!\cdot\!\Psi= \Psi^t\!\cdot\!(i\tau_y\mathcal{T})\hat{M}^t_{s}(i\tau_y\mathcal{T})\!\cdot\!(\Psi^\dagger)^t = -\Psi^t\!\cdot\!\hat{M}^{s}\!\cdot\!(\Psi^\dagger)^t\]
 to combine the Cooper and Fock terms. 

Now we assume that the there is some nonzero expectation value of the fields which corresponds to a non-zero order in one of the phases classified in the Appendix.
\begin{equation}
\langle\Psi\!\otimes\!\Psi^\dagger\rangle \equiv -\frac{1}{2Nc_\mathcal{A}}\left(\frac{m}{4\pi}\right) \sum_{\alpha}\hat{M}^{\alpha}_{\mathcal{A}}\Delta^{\alpha}_{\mathcal{A}},
\end{equation}
where matrices $\hat{M}^\alpha_\mathcal{A}$ are specified for each phase $\mathcal{A}$ in the Appendix and ${\bf \Delta_\mathcal{A}} = \{\Delta^\alpha_\mathcal{A}\}$ is the order parameter, which is singlet or multi-component depending on the phase. Below we will use the notation 
$|{\bf \Delta_\mathcal{A}}|^2\equiv\sum_\alpha|\Delta^\alpha_\mathcal{A}|^2$. The effective interaction constant 
$c_\mathcal{A}$ are defined for each phase as,
\begin{equation}
c_{\mathcal{A}}\equiv\sum_{s}g_{s}\left\{\delta_{s\mathcal{A}} - \frac{1}{4N^2}tr\left[\left(\hat{M}_s\hat{M}_{\mathcal{A}}\right)^{2}\right]\right\}.\label{eq:DefMFConstants}
\end{equation}
The assumption of a finite expectation value is consistent only if $c_\mathcal{A}<0$. The interaction mean field energy is therefore,
\begin{equation}
H_{sr} =\frac{1}{2}\sum_\alpha\Delta^\alpha_\mathcal{A}\left(\Psi^{\dagger}\!\cdot\!\hat{M}^\alpha_{\mathcal{A}}\!\cdot\!\Psi\right) - 
\frac{m}{8\pi c_\mathcal{A}}|{\bf \Delta}_\mathcal{A}|^2.
\end{equation}
Including the effect of the single-particle Hamiltonian $\hat{H}_0$, defined in Eq. (\ref{eq:H0Def}), the total mean field Hamiltonian is, 
\begin{equation}
H_{MF}\!=\!\frac{1}{2}\!\sum_k\Psi^\dagger(\boldsymbol{k})\!\!\left[\!\hat{H}_0\tau^N_z\!\!+\!\sum_\alpha\!\Delta^\alpha_\mathcal{A}\hat{M}^\alpha_\mathcal{A}\right]\!\!\Psi(\boldsymbol{k})
-\frac{m |{\bf \Delta_\mathcal{A}}|^2}{8\pi c_\mathcal{A}},
\label{eq:HMF}
\end{equation}
for fixed values of the order parameters $\Delta^\alpha_\mathcal{A}$. 

In the spirit of the Hartree-Fock or BCS theory, we diagonalize  Eq. (\ref{eq:HMF}) to obtain the ground state energy per unit area,
\begin{equation} 
E_{MF}\left(\{\Delta_\mathcal{A}\}\right) = -N\int_{\frac{k^2}{2m}< \mathcal{E}_c} 
\frac{d^2k}{2\pi}\varepsilon(k,{\bf \Delta_\mathcal{A}}) -\frac{m |{\bf \Delta_\mathcal{A}}|^2}{8\pi c_\mathcal{A}(\mathcal{E}_c)}.
\label{eq:MFEnergy}
\end{equation}
Here $\varepsilon(k,\{\Delta_\mathcal{A}\}))$ are the positive eigenvalues of the matrix $\hat{H}_0\tau^N_z\!+\!\sum_\alpha\!\Delta^\alpha_\mathcal{A}\hat{M}^\alpha_\mathcal{A}$ and the factor of $N$ comes from the degeneracy. The energy scale $\mathcal{E}_c$ is the energy scale at which we stop the RG.  The integral in Eq. (\ref{eq:MFEnergy}) evaluates to 
\begin{equation}
\int_{\frac{k^2}{2m}< \mathcal{E}_c} \dots =
{\rm const} + \frac{m|{\bf \Delta_\mathcal{A}}|^2}{2\pi}\left(\alpha_\mathcal{A} + \beta_\mathcal{A}\ln \frac{\mathcal{E}^2_c}{|{\bf \Delta_\mathcal{A}}|^2}\right),
\label{eq:MFRoughE}
\end{equation}
where $\alpha_\mathcal{A}$ and $\beta_\mathcal{A}$ are coefficients
 that depend on the phase $\mathcal{A}$.  The coefficients $\alpha_\mathcal{A}$ may be explicitly calculated from Eq. (\ref{eq:MFEnergy}). 
We list here $\alpha_\mathcal{A}$ for the states where $\Delta\cdot\ M$ may be written as $\sum_{ijkl}u_iv_jw_kx_l\tau^{AB}_i\tau^{KK'}_j\sigma_k\tau^{N}_l$.
Because of the 
 symmetry of $H_0$ there are only three independent coefficients. 
 Labeling the coefficient $\alpha$ by the representation and using 
the superscript $n$ or $s$ to denote normal or superconducting we have 
\begin{equation}
\begin{aligned}
\alpha_{A_1}^n&\!= \alpha_{A_2}^n\!=\alpha_{E_2''}^n \!=\alpha_{B_1}^s\!=\alpha_{B_2}^s\!=\alpha_{E_1''}^s = 1,\\
\alpha_{B_1}^n&\!= \alpha_{B_2}^n\!= \alpha_{E_1''} ^n \!=\alpha_{A_1}^s\!=\alpha_{A_2}^s\!=\alpha_{E_2''}^s=\!\frac{1}{2}\!+\!\log 2 \!\approx 1.19,\\
\alpha_{E_2}^n&\!=\alpha_{E_1}^n\!=\alpha_G^n= \alpha_{E_2}^s=\alpha_{E_1}^s=\alpha_G^s=\!\frac{1}{4}+\log 2\approx 0.94.\\
\end{aligned}
\end{equation}
The coefficients for the spin singlet and spin triplet normal states are the same because of symmetry. 
These coefficients are sufficiently close to one that we have simply taken $\alpha_\mathcal{A} \approx 1$. the term $\beta_\mathcal{A}\ln\frac{\mathcal{E}^2_c}{\Delta^2}$ in Eq. (\ref{eq:MFRoughE}) should be interpreted as the continuation of the RG flow from the scale $\mathcal{E}_c$ down to the energy $|{\bf \Delta}|$. Although obtained by using mean-field theory, since the flow of $c_\mathcal{A}$ is governed by the RG equation (\ref{eq:gRG}), we have to replace the logarithmic correction to Eq. (\ref{eq:MFRoughE}) with the evaluation of $c_\mathcal{A} $ at the energy $\Delta$. The mean field energy density is therefore written as,
\begin{equation}
E\left(\Delta_\mathcal{A}\right) = \frac{m}{8\pi}\left[-2N - \frac{1}{c_\mathcal{A}(|\Delta_\mathcal{A}|)}\right]|{\bf \Delta}_\mathcal{A}|^2,\  c_\mathcal{A}<0\label{eq:RedMFEnergy}
\end{equation}

 The ground state may now be determined by minimizing Eq. (\ref{eq:RedMFEnergy}) with respect to $\Delta_\mathcal{A}$ with the $c_\mathcal{A}(\Delta)$ obtained by numerical integration of the RG equations. 
The ground state energy gap will then be equal to the value of $|\Delta |$ at the minimum. 

It is important to note that we expect to find this minimum when the coefficient 
$c(|\Delta|)\sim \frac{1}{2N}$ which is inside of range of validity for our RG equation, $g \sim 1/N$.
The remaining subtlety is the inclusion of the long-range Coulomb interaction into
the mean-field description. Usually, it enters in the statically screened limit $g_{00z0}\simeq 1/N$ and does not
diverge at the transition.  
Furthermore, according to Eq.~(\ref{eq:DefMFConstants}), this constant can produce
only finite $1/N^2$ correction (positive to all supercoducting states and negative for all normal states).
Therefore, we will neglect $g_{00z0}$ in the further manipulations.

\section{The phase diagram}
\label{SEC5}
The result of minimizing Eq. (\ref{eq:RedMFEnergy}) is presented in Fig.~\ref{fig:bigPlots}. We find by extensive numerical investigation 
only five out of the possible sixty-four phases enumerated in Appendix
(10 in the charge channel, 22 in the spin and 32 in the Cooper channels). They are the nematic phase, the 
antiferromagnetic phase, the spin flux phase, and in the corners of the parameter space of bare interaction,  ferroelectric phase,
and singlet and triplet superconductor phases. 

Note that it is also possible that the resulting gaps are smaller than the energy of the Lifshitz transition\cite{Letter}, $\mathcal{E}_{LiTr} \approx 1meV$. In this case the renormalization of coupling constants is stopped at $\mathcal{E}_{LiTr}$, spontaneous symmetry breaking does not occur and the system remains in the symmetric state with the four Dirac cone spectrum.

Figure \ref{fig:bigPlots} shows the results of the RG analysis in terms of the resulting  symmetry broken phases. We find that there is significant variation in the scale $\mathcal{E}$, $\log(\mathcal{E}_0/\mathcal{E})\sim1\div20$, as expected from the wide range of couplings analyzed.
If we consider small initial couplings the $g_\mathcal{A}\ll1/N$ then the RG is driven by the constant term and $\log(\mathcal{E}_0/\mathcal{E})\sim10$ irrespective
of the initial conditions, resulting in a symmetry breaking only at extremely small energy scale $\mathcal{E} \sim 10^{-2}meV$. 

There is a variety of phases that have been proposed as the ground state of BLG that we do not find. An anomalous quantum Hall state (QAH) state was suggested in 
Ref.~\onlinecite{Levitov2}, corresponding to the representation $g_{A_2}$ with order parameter $\langle\psi^\dagger\tau^{AB}_z\psi\rangle$.
This is not found as a ground state in our analysis.  In the same paper, and in Refs.~\onlinecite{MacDonald2, MacDonald}, a large manifold of quantum hall ferromagnetic states were suggested containing the representations $E_2''$, $A_2$ and $B_2$ in both spin singlet and spin triplet representation.
 All of these states were considered as degenerate appealing to the $SU(4)$ symmetry of the single particle Hamiltonian, Eq. (\ref{eq:H0Def}).  
In both cases the artificial  $SU(4)$ symmetry was assumed to be exact, which is contradicted by the importance of the short range interactions we find here in solving the complete set of RG equations. 

\begin{figure}
\includegraphics[width =0.9\columnwidth]{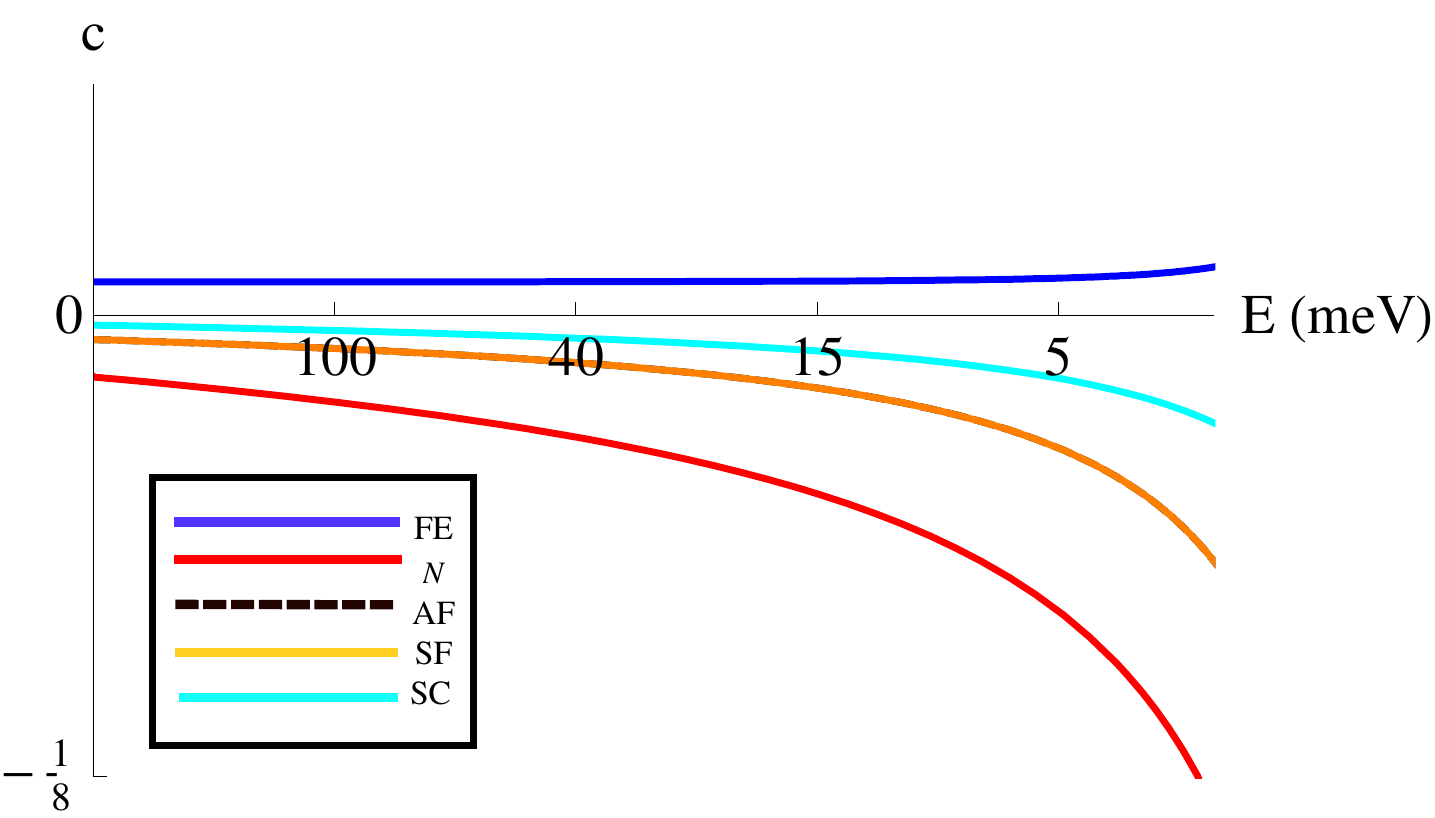}
\caption{ (Color online)Interaction energy as a function of energy scale for selected phase. For these initial conditions the nematic phase is the ground state}\label{fgi:NPhase}
\end{figure}

In subsections below, we discuss the details of each phase. We will present a comparative flows of the couplings
defined in Eq.~(\ref{eq:DefMFConstants}) to illustrate the competition between phases, see Figs.~\ref{fgi:NPhase} -- \ref{Fig15}.

\subsection{Nematic Phase (N)}

In the nematic phase, there is a finite expectation value for the order parameter $\langle\psi^{\dagger}\tau^{KK'}_{z}\tau^{AB}_{x,y}\psi\rangle$,  breaking the rotational symmetry of the system from six-fold to two-fold while maintaining translational symmetry.  
The order parameter is characterized by enhanced electron hopping in one direction.
The interaction energy for this phase depends on the combination of parameters obtained from Eq.~(\ref{eq:DefMFConstants}):
\begin{equation}
c_N = \frac{1}{8 }g_{B_2} - \frac{1}{4}g_{E_2''} +g_{E_2}  + \frac{1}{8}g_{B_1} - \frac{1}{8} g_{E_1''} - \frac{1}{8} g_{A_2}.
\label{eq:NEnergy}
\end{equation}

It is the preferred ground state in the absence of intervalley scattering, and it is also generally the ground state when the bare $g_{E_2}$ coupling is negative. Note that a negative contribution towards bare $g_{E_2}$ comes from the electron-electron interaction via the polarization of the lattice, ie. via virtual excitation/absorption of in plane phonons near the $\Gamma$ point. The nematic phase is also the ground state over other large parts of the parameter space as can be seen in Fig. \ref{fig:bigPlots}. This reflects the fact that the coupling $g_{E_2}$ almost always becomes negative rapidly (see Fig. \ref{fig:RGRunning}).

We previously proposed the nematic phase as a possible ground state in based on a more limited analysis of the RG equations \cite{Letter}. Using a similar renormalization group analysis Vafek and Yang\cite{Vafek,Vafek2} also find the nematic phase as a possible ground state, supporting the analysis in this paper.  

The most notable characteristic of the nematic phase is that it remains gapless, but with the parabolic bands reconstructed into two Dirac mini-cones at energies less than $|{\bf\Delta}|$. The nematic phase would show metallic behavior in conductance measurements, but decreasing density of states at low densities.  The nematic 
state preserves time reversal symmetry and so it should not show asymmetry between positive and negative magnetic fields in magneto-transport. The nematic order parameter transforms in the same representation of $\mathcal{D}_{3d''}$ as uniaxial strain so that strain will couple directly to this state.  It is possible that strain could induce a transition into the nematic phase\cite{NewStrain}, even if unperturbed BLG chooses another phase as the ground state.
 
\subsection{Anti-ferromagnetic Phase (AF)}

The anti-ferromagnetic phase is defined by a nonzero expectation value of $\langle\psi^\dagger\tau^{AB}_z\tau^{KK'}_z\vec{\sigma}\psi\rangle$. This state corresponds to opposing magnetic moments on the $A$ and $B$ sublattices. The orbital part breaks the reflection symmetry between the two sublattices but otherwise preserves the $\mathcal{D}_{3d}''$ symmetry of the BLG in its entirety. 

The exchange energy depends on the following combination of coupling constants  obtained from Eq.~(\ref{eq:DefMFConstants}):
\begin{equation}
\begin{split}
c_{AF} & = -\frac{1}{2}g_G -\frac{1}{8}g_{B_2} 
\\
& + \frac{1}{4}\left(g_{E_2''} +2g_{E_2} + g_{E_1}  + g_{E_1''} \right)- \frac{1}{8}g_{B_1} - \frac{1}{8}g_{A_2}
\end{split}
\label{eq:AFEnergy}.
\end{equation}
The AF promoted strongly by the coupling $g_{G}$, with the factor of four coming from the dimension of the representation $G$. This effect is amplified by the sensitivity of the RG equations to the coupling $g_G$. As a result even small values of $g_G$ near the free field point make the AF state the ground state. The AF state is also promoted by negative $g_{E_2''}$  and is generally the ground state when we start with negative $g_{E_2''}$. Again we emphasize that these conclusions come from the combination of RG equations and interaction energy, not just from the interaction energy alone. 
\begin{figure}
\includegraphics[width=0.9\columnwidth]{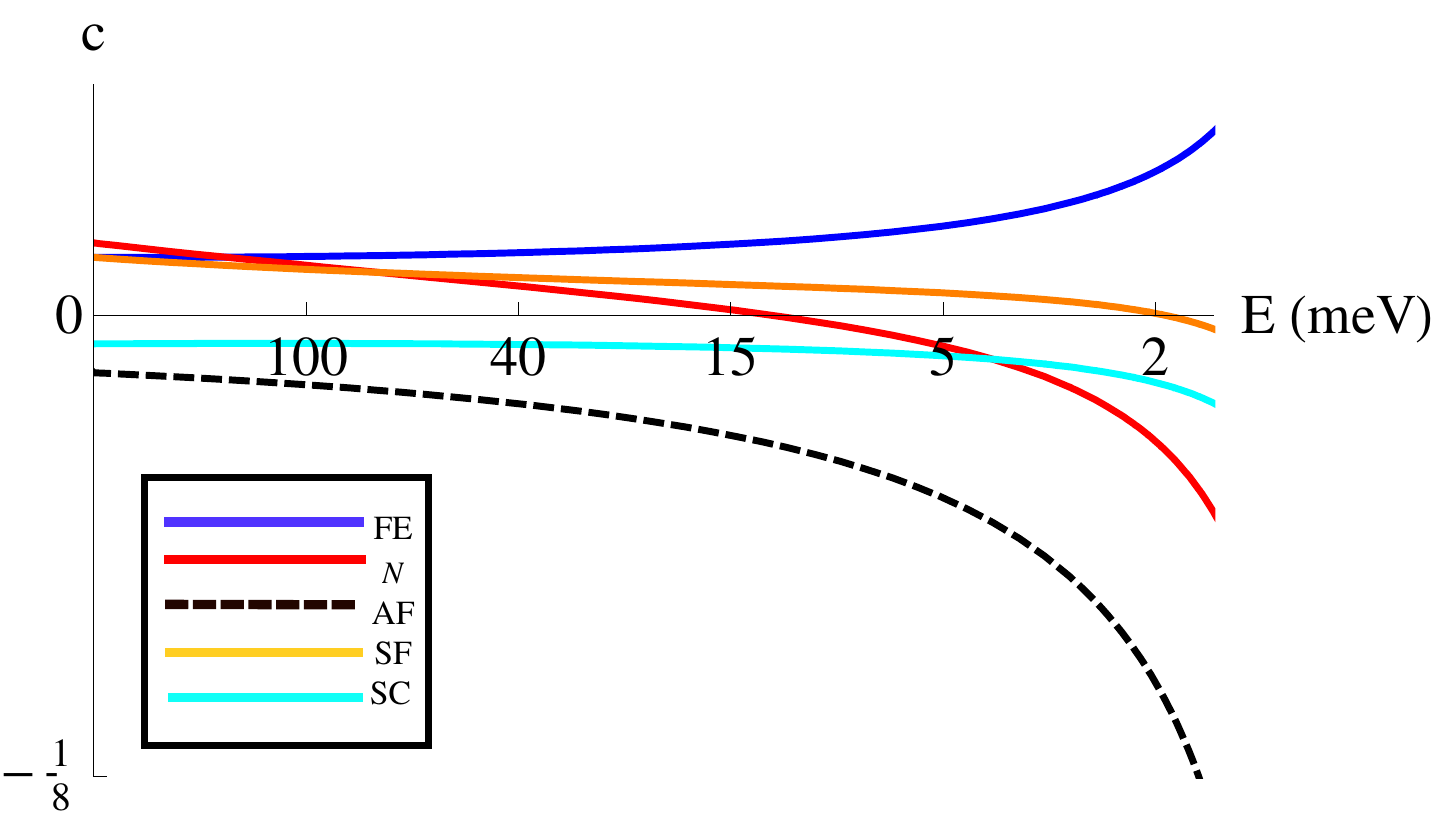}
\caption{ (Color online)The interaction energy as a function of energy scale for some choice of initial parameters. In this case the ground state is the AF phase.}
\end{figure}
Other authors have proposed this AF state as a possible ground state. Kharitonov \cite{Kharitonov} suggested the AF state based on experimental evidence and simple mean field theory arguments applied at the high-energy scale directly. The structure of our RG equations indicate that such a simple mean field theory is not applicable to BLG, although it does suggest the same state. 

The AF phase is expected to be an insulating state with activated gap behavior in transport measurements. Although the magnetic field does not couple directly to the AF state since it is antiferromagnetic, the Zeeman energy splitting does break the $SU(2)$ spin symmetry of the system (spin-orbit is negligible). Therefore the gap should show a decline in tilted magnetic field.  The LAF state is adiabatically connected to a quantum Hall ferromagnetic state at higher magnetic field. A lack of features between the zero and high magnetic field state might be the 
evidence that the zero magnetic field state is AF\cite{Kharitonov}.

\subsection{Spin flux (spin Hall) Phase (SF)}

The spin flux (SF) phase (elsewhere called a quantum spin Hall state) is defined by the finite expectation value of $\langle\psi^\dagger\tau^{AB}_z\vec{\sigma}\psi\rangle$. The effect of this on the electrons is equivalent to the development of a finite spin orbit coupling, and gaps the electronic spectrum. It may be viewed as a state where spin currents circle the honeycomb rings, or as a quantum anomalous Hall effect state, but with opposite signs for opposite spins, producing no net charge current, so that this state preserves time reversal invariance.

The interaction energy of the SF depends on the combination of coupling constants  obtained from Eq.~(\ref{eq:DefMFConstants}):
\begin{equation}
c_{SF} =\frac{1}{4}\left(g_{E_2''} - g_{E_2}  + g_{E_1}  - g_{E_1''}\right) -\frac{1}{8}
\left(g_{B_2} + g_{B_1}+g_{A_2}\right) .
\label{eq:SFEnergy}
\end{equation}
\begin{figure}
\includegraphics[width =0.9\columnwidth]{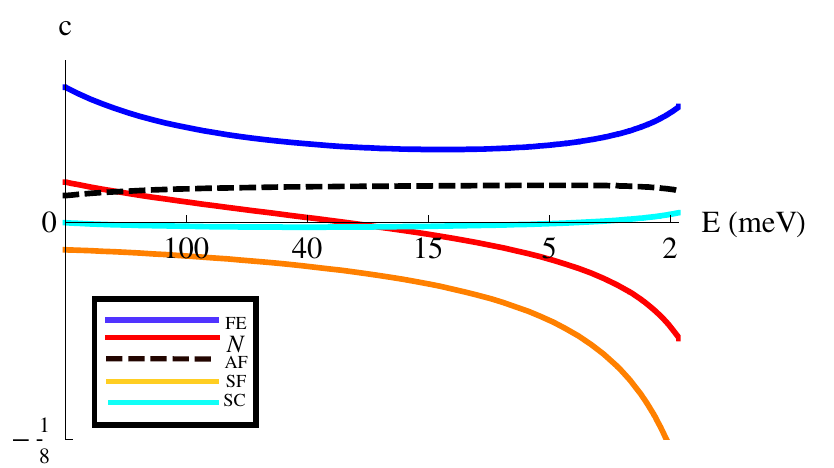}
\caption{ (Color online)Interaction energy for selected phases a function of energy scale. In this case the ground state is the spin flux state shown in orange}
\end{figure}
The SF  state has a similar matrix structure as the AF state and therefore a similar interaction energy. However it is not promoted by large $g_G$, unlike the AF phase. Therefore, generically, large $g_G$ generally suppresses the SF in favor of other states as can be seen in Fig. \ref{fig:bigPlots}.

Analogously to the case of spin-orbit coupling in monolayer graphene\cite{KaneMele}, the finite value of the spin flux OP may create a "spin Hall effect" with quantized spin Hall conducivity.  The edge states 
and insulating bulk imply a quantized conductance of $4e^2/h$, unless they are localized by magnetic and intervalley-scattering disorder. 
The state is time reversal invariant, so no transverse conductance at zero magnetic field is possible.

\subsection{Limits of applicability: Ferroelectric Phase}

The ferroelectric (FE) phase is characterized by a non-zero expectation value of $\langle\psi^\dagger\tau^{AB}_z\tau^{KK'}_z\psi\rangle$. It is a spontaneous charging of the BLG with opposite charge on the two layers. 

The interaction energy for this phase depends on the combination of couplings  obtained from Eq.~(\ref{eq:DefMFConstants}):
\begin{equation}
c_F= \frac{7}{8}g_{B_2}  + \frac{1}{4}\left(g_{E_2''} - g_{E_2}  + g_{E_1}  - g_{E_1''}\right) - \frac{1}{8}g_{B_1}- 
\frac{1}{8}
g_{A_2}.
\end{equation}
We find that the ferroelectric phase is strongly suppressed by positive bare $g_{B_2}$ and the development of the FE phase requires the $g_{B_2}$ coupling to diverge to negative infinity, which can only happen in a small sliver of the phase space where the combination of the higher order diagrams conspire to drive $g_{B_2}$ negative. Even in this section the energy difference between the ferroelectric and competing phases is never large, and it may be that in a more accurate calculation it never appears as the ground state.
\begin{figure}
\includegraphics[width =0.9\columnwidth]{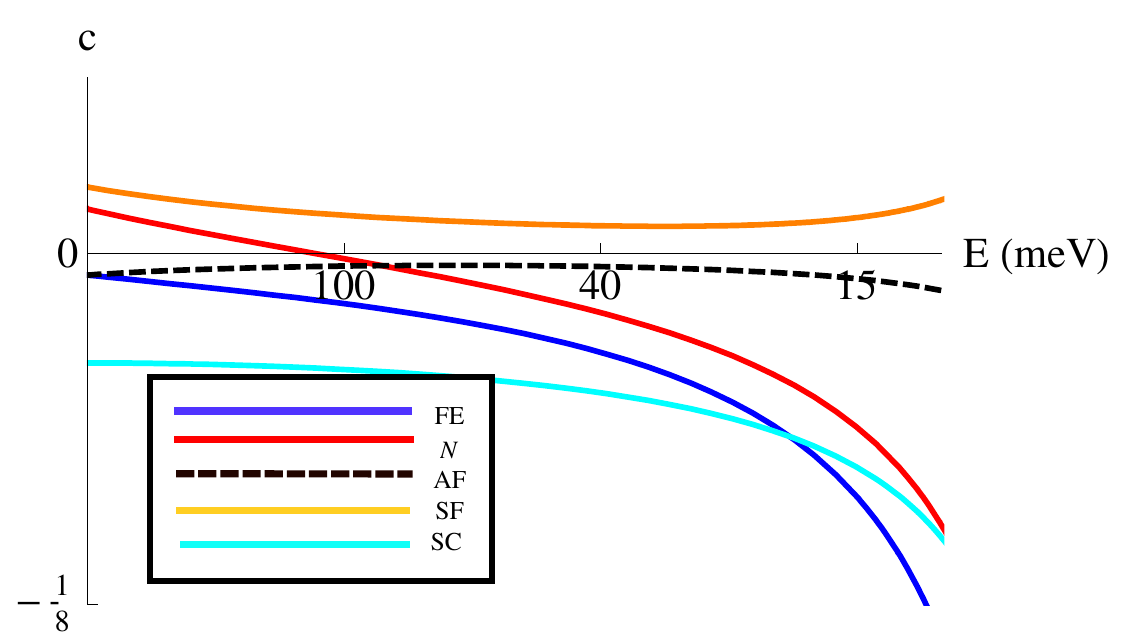}
\caption{ (Color online)The evolution of the interaction energy of the selected phases as a function of the energy scale. In this case the ground state is a ferroelectric phase. The difference in energy between the F phase and nearby phases is quite small. }
\end{figure}

The FE state was also proposed in Ref.~\onlinecite{Levitov}. However this analysis was based on a flawed mean field theory treatment counting only the diagrams from 
Fig.~ \ref{fig:FeynRG} (c) marked (i) and (ii) as well as considering only the single parameter scaling theory with one interaction constant in the $B_2$ channel. 
These are not parametrically justified approximations.

The ferroelectric phase is a completely gapped state, with neither neutral nor charged excitations. It is also a trivial insulator in that it does not possess protected edge modes. 
Therefore, it should display insulating transport behavior. An external field perpendicular to the BLG flake would promote the FE phase, increasing the gap due to the interlayer symmetry breaking at the single particle level \cite{McCann}. This does not seem to take place in any of the recent experiments \cite{Weitz,Velasco} on BLG, where the external transverse field destroyed the zero-field state, and introduced a distinct state determined by the interlayer asymmetry.

\subsection{Limits of applicability: Superconducting Phases}

The singlet superconducting (SS) phase is characterized by the usual order parameter 
$\langle \psi^\dagger  \mathcal{T}\psi^\dagger\rangle$.
The coupling for this phase is found  from Eq.~(\ref{eq:DefMFConstants}) to be:
\begin{equation}
\begin{split}
c_{SS}&=\frac{1}{2}g_G + \frac{1}{8}g_{B_2}
\\ & + \frac{1}{4}\left(g_{E_2} + g_{E_2''}  - g_{E_1} - g_{E_1''}\right)  - \frac{1}{8}g_{A_2} - \frac{1}{8}g_{B_1}.
\end{split}
\end{equation}
The stability of this phase requires very significant negative couplings from the very beginning
and the phase can not arise from the purely repulsive interaction. 
\begin{figure}
\includegraphics[width=0.9\columnwidth]{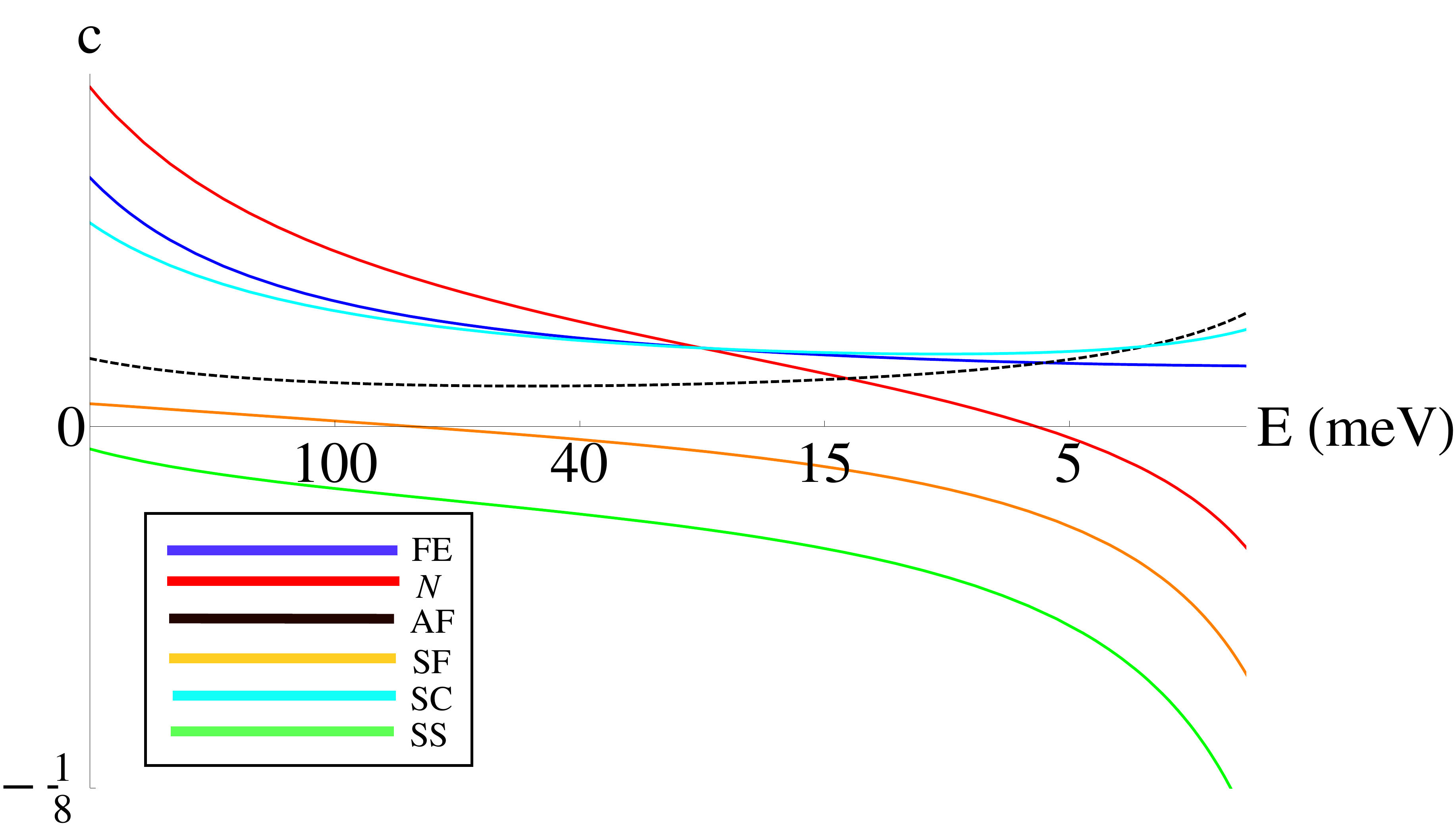}

\caption{ (Color online)Interaction energy for selected phases as a function 
of the energy scale. In this case the singlet superconducting phase is the ground phase.}
\end{figure}

The triplet superconducting (SC) phase has  the order parameter 
$\langle \psi^\dagger \tau_z^{KK'}\vec{\sigma} \mathcal{T}\psi^\dagger\rangle$, with 
the pairing function of the opposite signs in the $K$ and $K'$ valleys. 

 Its interaction depends on the combination of couplings  obtained from Eq.~(\ref{eq:DefMFConstants}):
\begin{equation}
\begin{split}
c_{SC} & = -\frac{1}{2}g_G+\frac{1}{8}g_{B_2}\\ & +
\frac{1}{4}\left( g_{E_2} -g_{E_2''}  - g_{E_1} +g_{E_1''} \right) - \frac{1}{8} g_{A_2}- \frac{1}{8} g_{B_1}.
\end{split}
\label{Eq:SCEnergy}
\end{equation}
\begin{figure}
\includegraphics[width=0.9\columnwidth]{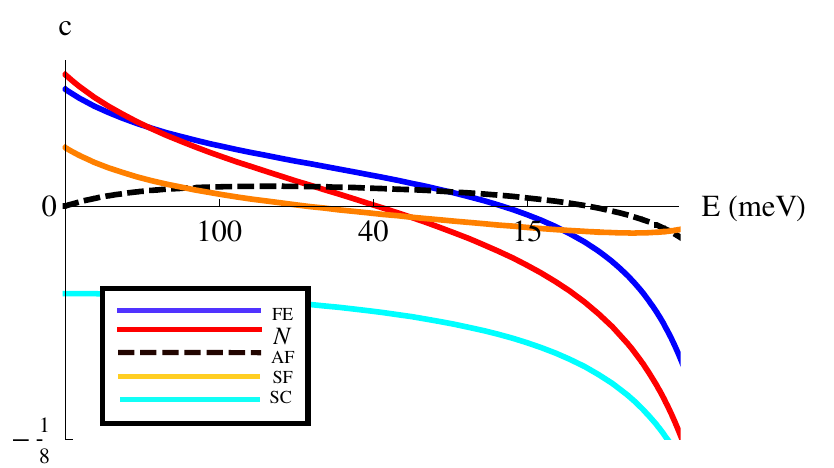}

\caption{ (Color online)Interaction energy for selected phases as a function 
of the energy scale for initial repulsive short range interaction. In this case the triplet superconducting phase is very close to the ground phase.
Even though $c_{SC}$ appears to be the most negative, minimization  of Eq.~(\ref{eq:RedMFEnergy}) gives
the nematic state as a preferd ground state.
}\label{Fig15}
\end{figure}
For the repulsive interaaction, the triplet SC phase only appears as a stable phase at large couplings.

To conclude, both superconducting phases appear only on the limits of the applicablity of the theory.
Moreover, finite value of the underscreened Coulomb interaction push the energies of those stated further up.
It is therefore possible that the appearance of the SC phases is merely an artifact of our 
inability to deal properly with strong couplings, and that the superconductivity
does not belong to  the actual phase diagram.

\section{Conclusion}

Several results have been established in this paper.  The ground state of BLG cannot be understood without considering the high energy couplings in detail, because different couplings lead to different ground states. Moreover naive expectations about the importance of certain couplings are not borne out and all plausible combinations must be considered. In particular excluding intervalley scattering leads to misleading results. The previously reported nematic state is the ground state for a significant fraction of these couplings. Of the large number of ground states that are possible we find only five  appear. They are nematic, antiferromagnetic, ferroelectric and triplet superconductor, as well as a "spin flux" phase not previously proposed.  The nematic, antiferromagnetic and spin flux phases seem the most likely candidates.  

The present work may be extended in a variety of ways. The accuracy of the renormalization group equations may be improved and validated by considering higher order diagrams and a more detailed mean field theory constructed. One may also try to connect the value of the couplings at the scale $\gamma_1/2 \approx 0.2meV$ with their value at the bandwidth of the $\pi$ orbitals. The possible phase transitions between the proposed states and the behavior of domain walls between regions of different phases may be needed to properly account for the transport data.

 Such improvements aside, the unique challenge of theoretically determining  the electronic ground state of BLG has been laid out. The problem naturally involves the competition of an uncommonly large set of phases and interactions. Truncating the theory to a more tractable subset does not appear to give accurate results. Instead the problem must be attacked in its full complexity.

\acknowledgements

This work was supported by the ERC Advanced Investigator Grant, EPSRC grant EP/G041954, and Royal Society. 
Also, the authors thank 
M. Kharitonov for reading the manuscript,  M. Mucha-Kruczynski for technical assistance, and 
KITP-UCSB Research Programme ``The Physics of Graphene'' for hospitality during the final stage of this work.

\appendix*
\section{Group theory for phases in BLG}
\label{Appendix}
In this appendix we classify the possible phases of BLG. We will use the matrix notation defined in Sec.~\ref{SEC2}.

Phases are defined by all possible expectations $\Delta \equiv \langle\Psi\otimes \Psi^\dagger\rangle$ 
that belong to an irreducible representations (irrep) of the symmetry group ${\cal G}$ of BLG. 
Every phase defines a subgroup ${\cal H}$ of ${\cal G}$ consisting of all operations that leave $\Delta$ invariant.  
Two phases within an irrep are distinct if their invariant subgroups are not conjugate. 
(Recall two subgroups $\cal{H}$ and $\cal{H}'$ of a group ${\cal G}$ are conjugate if they is an element of $g \in {\cal G}$ such that $g{\cal H}g^{-1} = {\cal H}'$). 
This definition is correct is in the sense that it gives all physically distinct states that may be reached via a second order phase transition at the 
highest critical temperature, per the usual Landau theory. 

\begin{figure}
\includegraphics[width=0.9\columnwidth]{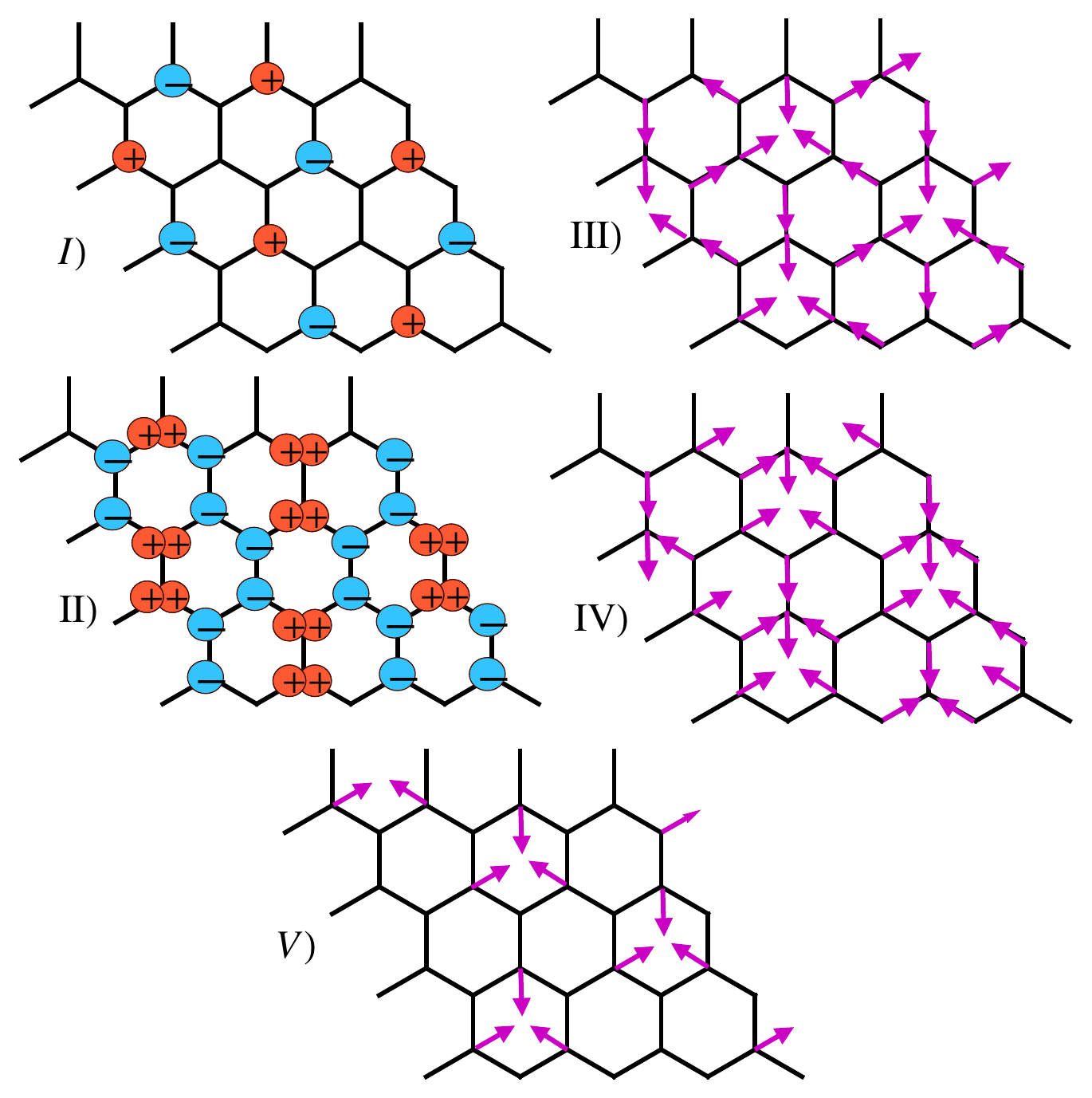}
\caption{ (Color online)Sketch of the symmetric phases magnetic and normal phases transforming according the G representation.
(I) $G_1$ normal state; (II) $G_2$ normal state; (III) $G_3$ spin state; (IV) $G_4$ spin state; (V) $G_5$ spin state. 
 Because of the absence of spin-orbit coupling the overall direction of the spins is arbitrary.}
\label{fig:GPhasesFig}
\end{figure}

Let us notice, however, that the anomalous averages belonging to
 the same irrep of the original  group ${\cal G}$ may correspond to the different phases.
For example, $\Delta_I = \tau^{AB}_y\tau^{KK'}_y$ and $\Delta_{II} = \tau^{AB}_x\tau^{KK'}_x+\tau^{AB}_y\tau^{KK'}_y$ are both charge density waves 
with a tripled unit cell transforming in the $G$ representation.
However, these two phases are distinct since $\Delta_{II}$ is invariant under 
rotations by $2\pi/3$ around a lattice site, whereas $\Delta_I$ is not invariant 
under any conjugate operation, see Fig. \ref{fig:GPhasesFig}. As a further example, a canted anti-ferromagnetic phase would be 
given by $\Delta = \tau^{AB}_x\tau^{KK'}_z\sigma_x  + \sigma_z$.
This is not considered in the present classification since it does not belong to an irrep of ${\cal G}$ (it is a linear combination elements of the $B_2$ and $A_1$ representations).  
Of course all such mixed states may be constructed from linear combinations of phases in our classification.

We classify the phases according to the symmetry group ${\cal G}$ of BLG which is 
effective at intermediate energies, see Sec.~\ref{SEC2}. 
In this regime the effect of RG irrelevant perturbations such as Umklapp scattering may be ignored, and RG relevant but weak perturbations, such as trigonal warping and spin-orbit coupling, may be neglected. This approximation should be effective at energies between $\gamma_1/2 = 0.2eV$ and $\mathcal{E}_{LiTr} = 1meV$, 
which contains any energy scale associated with spontaneous symmetry breaking. In this regime the symmetry group is 
\begin{equation}
{\cal G} = \mathcal{D}^{(rot)}_{\inf}\times\mathcal{D}^{(tran)}_{\inf}\times SU(2)^{(spin)}\times U(1)^{(gauge})\times T. 
\label{eq:DefineG}
\end{equation}
The first subgroup, $\mathcal{D}^{(rot)}_{\inf}$, is generated by infinitesimal spatial rotations $C$ and inversion $R_C$ of the BLG plane.  
The second subgroup, $\mathcal{D}^{(tran)}_{\inf}$, is generated by an infinitesimal translation $t$  and reflection $R_t$. 
The action of these operators on the low-energy electrons is given in terms of the Pauli matrices:
\begin{equation}
\begin{aligned}
C \equiv& i\tau^{AB}_z\\\
t \equiv & i\tau^{KK'}_z\\
R_C \equiv& \tau^{AB}_{x}\tau^{KK'}_z\\
R_t \equiv& \tau^{AB}_{z}\tau^{KK'}_x\\
R_t\cdot R_v = & \tau^{AB}_{y}\tau^{KK'}_y\\
\end{aligned}
\label{def:DefineGGen}
\end{equation}
Note that these two groups commute with each other, unlike true translation and reflection.  
In the presence of the appropriate symmetry breaking these are reduced down to the $D_{3d''}$ group discussed in the text.
In this case the continuous translations and rotations $\exp\left(\theta_tt +\theta_c C \right)$ become discrete with $\theta_{t,c}=0,\pm 2\pi/3$ 
and the inversions $R_t$ and $R_C$ become $R_h\cdot R_v$ and $R_v$ of Sec.~\ref{SEC2} respectively.

The $SU(2)^{(spin)}$ is the group of spin rotations, which is decoupled from the physical rotations because there are no spin-orbit interactions. It is generated by rotations $\vec{S}=(S_x,S_y,S_z)$. The high symmetry axis of any phase will always chosen to be z. 
Infinitesimal rotations around the $z$ axis are represented by $S_z$. 
There are also reflections and inversion of the spin space, but these do not distinguish any phases, so we will suppress them.

The gauge groups acts by multiplication. We label the infinitesimal generator $g$, which acts on the wavefunctions $\psi$ simply by $g\psi = i\psi$. 

The time reversal operator $T$ commutes with all of the above except the gauge generator $TgT = -g$. It is given in terms of Pauli matrices by 
\begin{equation}
T\equiv i\sigma_y\tau^{KK'}_y\tau^{AB}_y K,
\label{eq:reDefT}
\end{equation}
where $K$ is complex conjugation.

This defines the symmetry group ${\cal G}$ fully.  A further enlarged symmetry group isomorphic to $U(1)\times U(4)$ is 
considered at points in the body of the text [see Eq. (\ref{eq:SU(4)gen})] and other works, but there is no reason to expect that this will every be an accurate approximation. We will proceed to categorize the phases according to the symmetry group ${\cal G}$. 
One may always collapse classification to obtain the distinct phases under the artificially enlarged symmetry groups.

In the following subscections we present the tables enumerating the possible symmetry breaking
for the singlet, triplet, and superconducting phases each. Each table is split into 
subsections corresponding to the IrReps of ${\cal D}_{3d''}$.
 These are listed in the first column. At the beginning of each subsection the second column gives the order parameter (OP) 
for the IrRep in terms of the notation $\hat{\Delta}\equiv\Delta_{abc}\tau^{AB}_a\tau^{KK'}_b\sigma_c$ and arranges these into a vector. 
Next to this are the generators of symmetries under which the order parameter is invariant (generators and $\Delta$ commute). 
This completely characterizes one dimensional represntations.
 
In the case of the multi-component representations, particular values of the order parameter may have higher symmetries than the generic values - these are the distinct phases. 
The values of the order parameter that produce the phase are given according to the vector representation in the second column. 
Next to these in the third column are the additional residual symmetries under which the phase is invariant, and the phase is labeled with the additional subscript.

  For example, let us take in section $G$ in the first table. The second column of first line defines a vector for the representation. 
The third column states that all vectors in that representation are invariant under the time $T$ time reversal operation.
The next line says that when the vector is proportional to $(1,0,0,1)$, i.e. $\Delta \propto \tau^{AB}_x\tau^{KK'}_x + \tau^{AB}_y\tau^{KK'}_y$ , the symmetry is higher. 
The higher symmetries are in the third column; in particular, this vector is invariant under the combined $C+t$ rotation and the $R_C\cdot R_t$
 reflection in addition to the $T$ rotation.
 The next line of the table says that when the vector takes the value $(0,0,0,1)$ the state is invariant under the reflections $R_t, R_C$. 
Note that each phase is generally defined by a coset of values of the order parameter, which are all invariant under conjugate groups. 
We only list one representative from each coset. For example, in the case of $G$, the $(1,0,0,1)$ vector is part of the coset of vectors $\left(\cos\theta, \sin\theta,-\sin\theta,\pm\cos\theta\right)$, $\theta \in [0,2\pi]$. 
These are all invariant under subgroups conjugate to the one listed in the third column.  

We use $\alpha$, $\beta$ as arbitrary real parameters when there is a continuous manifold of cosets. When listed under symmetries the symbols 
$t$, $C$, $g$ and $S_z$ mean the phase is invariant under the entire $U(1)$ group generated. The symbol $\vec{S}$ means the phase is invariant under all spin rotations. There are several symmetry operations involving rotations by $\pi$ or $\pi/2$ in one the $U(1)$ groups. Like the spin reflections, these do not distinguish any of the phases so we do not list them. 
Five of the phases belonging to the $G$ representation are illustrated in Fig.~\ref{fig:GPhasesFig}.
Notice that, according to the Landau theory, the  transition to the $G$ -type phase can not occur directly but rather through
the pattern with incommensurate periodicity.  

%The corresponding lines in the tables are marked with Roman numerals.

\subsection{Normal Phases}
The normal phases are by definition invariant under spin and gauge transformation so we will suppress them.
A product of Pauli matrices acting in different sub-spaces should be understood as a direct product. 

\begin{longtable}{|p{.4cm}|p{3.5cm}|p{2.2cm}|p{2.1cm}|}
\hline\hline
Irr. & OP & $M^\alpha$ & Symmetry\\
\hline \hline
$A_2$ & $\Delta_{0z0}$ & $\tau_z^{AB}\openone^{KK'}\openone^N \openone^s$ & $ C, t, R_C$ \\
\hline\hline
$B_1$ & $\Delta_{z00}$ & $\openone^{AB}\tau_z^{KK'}\openone^N \openone^s$  &$C, t, R_t$\\
\hline\hline
$B_2$ & $\Delta_{zz0}$ &$\tau_z^{AB}\tau_z^{KK'}\tau_z^N \openone^s$& $C, t, R_C\cdot R_t,T$\\
\hline\hline
$E_2$ & $\left(\Delta_{xz0}, \Delta_{yz0}\right)$& $\tau_{x,y}^{AB}\tau_z^{KK'}\tau_z^N \openone^s$  & $t, R_C, R_t, T$\\
\hline\hline
$E_2''$ & $\left(\Delta_{zx0}, \Delta_{zx0}\right)$& $\tau_{z}^{AB}\tau_{x,y}^{KK'}\tau_z^N \openone^s$& $C, R_C, R_t, T$\\
\hline\hline
$E_1$ & $\left(\Delta_{x00},\Delta_{y00}\right)$ & $\tau_{x,y}^{AB}\openone_z^{KK'}\openone^N \openone^s$ &$t, R_C,R_t\cdot T$\\
\hline\hline
$E_1''$ & $\left(\Delta_{0x0}, \Delta_{0y0}\right)$& $\openone^{AB}\tau_{x,y}^{KK'}\openone^N \openone^s$ & $C,R_C, R_t\cdot T$\\
\hline\hline
$G$ & $\left(\Delta_{xx0}, \Delta_{xy0}, \Delta_{yx0}, \Delta_{yy0}\right)$ &
 $\tau_{x,y}^{AB}\tau_{x,y}^{KK'}\tau_z^N \openone^s$
& $T$ \\
\hline 
   $G_1$	& $\left(1,0,0,1\right)$  && $ C+t, R_C\cdot R_t$ \\
\hline
   $G_2$	& $\left(0,0,0,1\right)$  & &$ R_t, R_C$ \\
\hline\hline
\end{longtable}

\subsection{Magnetic phases}
We restore the spin symmetries but continue to suppress the gauge symmetry. The high symmetry axis is arbitrarily chosen to be the z direction.

\begin{longtable}{|p{.55cm}|p{2.7cm}|p{2.65cm}|p{2.2cm}|}
\hline\hline
Irr. & OP &$M^\alpha$  & Symmetry\\
\hline \hline
$A_1$ & $\left(\Delta_{00x},\Delta_{00y},\Delta_{00z}\right)$ & 
 $\openone^{AB}\openone^{KK'}\openone_z^N \sigma_{x,y,z}$
&$ C, t, S_z, R_t, R_C$ \\
\hline \hline
$A_2$ & $\left(\Delta_{0zx},\Delta_{0zy},\Delta_{0zz}\right)$ &
 $\openone^{AB}\tau_z^{KK'}\tau_z^N \sigma_{x,y,z}$
& $ C, t, S_z, R_C, T$ \\
\hline\hline
$B_1$ & $\left(\Delta_{z0x},\Delta_{z0y},\Delta_{z0z}\right)$ &
 $\tau_z^{AB}\openone^{KK'}\tau_z^N \sigma_{x,y,z}$
& $C, t, S_z, R_t, T $\\ 
\hline\hline
$B_2$ & $\left(\Delta_{zzx},\Delta_{zzy},\Delta_{zzz}\right)$ &
 $\tau_z^{AB}\tau_z^{KK'}\openone^N \sigma_{x,y,z}$
& $C, t, S_z,\quad\linebreak R_C\cdot T, R_t \cdot T\quad $\\
\hline\hline
$E_2$ & $\left(\!\Delta_{xzx},\Delta_{xzy},\Delta_{xzz};\right.\linebreak 
\left.\Delta_{yzx},\Delta_{yzy},\Delta_{yzz}\!\right)$& 
$\tau_{x,y}^{AB}\tau_z^{KK'}\openone^N \sigma_{x,y,z}$
& $t, R_C,R_t$\\
\hline
$E_{2,1}$
	& $\left(0,0,1;0,0,0\right)$  & &$S_z$\\
\hline
$E_{2,2}$	& $\left(1,0,0;0,1,0\right) $ & & $S_z+C$\\
\hline\hline
$E_2''$ & $\left(\!\Delta_{zxx},\Delta_{zxy},\Delta_{zxz};\right.\linebreak\left.
\Delta_{zyx},\Delta_{zyy},\Delta_{zyz}\!\right)$ & 
$\tau_{z}^{AB}\tau_{x,y}^{KK'}\openone^N \sigma_{x,y,z}$
& $C, R_C,R_t$\\
\hline
$E_{2,1}''$	& $\left(0,0,1;0,0,0\right)$  && $S_z$\\
\hline
$E_{2,2}''$	& $\left(1,0,0;0,1,0\right) $  && $S_z+t$\\
\hline\hline
$E_1$ & $\left(\!\Delta_{x0x},\Delta_{x0y},\Delta_{x0z};
\right.\linebreak\left.
\Delta_{y0x},\Delta_{y0y},\Delta_{y0z}\!\right)$ &$\tau_{x,y}^{AB}\openone^{KK'}\tau_z^N \sigma_{x,y,z}$ & $t,R_t, T$\\
\hline
$E_{1,1}$
	& $\left(0,0,1;0,0,0\right)$  & &$S_z$\\
\hline
$E_{1,2}$
	& $\left(1,0,0;0,1,0\right) $  & &$S_z+C$\\
\hline\hline
$E_1''$ & $\left(\!\Delta_{0xx},\Delta_{0xy},\Delta_{0xz};\right.\linebreak\left. 
\Delta_{0yx},\Delta_{0yy},\Delta_{0yz}\!\right)$ 
&$\openone^{AB}\tau_{x,y}^{KK'}\tau_z^N \sigma_{x,y,z}$ 
& $C,R_C, T$\\
\hline
$E_{1,1}''$	& $\left(0,0,1;0,0,0\right)$ & & $S_z$\\
\hline
$E_{1,2}''$	& $\left(1,0,0;0,1,0\right) $ & & $S_z+t$\\
\hline\hline
$G$ & $(\!\Delta_{xxx},\Delta_{xxy},\Delta_{xxz}\!;
\linebreak
\Delta_{xyx},\Delta_{xyy},\Delta_{xyz}\!;
\linebreak\Delta_{yxx},\Delta_{yxy},\Delta_{yxz};\linebreak \Delta_{yyx},\Delta_{yyy},\Delta_{yyz}\!)$ &
$\tau_{x,y}^{AB}\tau_{x,y}^{KK'}\openone^N \sigma_{x,y,z}$ 
& None \\
\hline
$G_1$   & $\left(0,0,1;0,0,0; \right.\linebreak\left. 0,0,0;0,0,1\right)$ & & $S_z, C + t,R_C\cdot R_t$ \\
\hline
 $G_2$ 	& $\left(0,0,0;0,0,0;\right.\linebreak\left. 0,0,0;0,0,1\right)$ & & $S_z,R_C,R_t$ \\
\hline
 $G_3$  & $\left(0,0,0;1,0,0;\right.\linebreak\left. 0,0,0;0,1,0\right)$  & & $S_z+t, R_C$ \\
 \hline
 $G_4$  	& $\left(0,0,0;0,0,0;\right.\linebreak\left. 1,0,0;0,1,0\right)$  & & $S_z+ C, R_t$\\
 \hline
  $G_5$  	& $\left(1,0,0;0,1,0;\right.\linebreak\left. 0,-1,0;1,0,0\right)$  & & $S_z+C, S_z+t$\\
\hline\hline
\end{longtable}

\vspace*{1cm}

\subsection{Superconducting Phases}
The order parameter $M$ is defined by the non-zero expectation values $\langle\psi^\dagger {M}{T} \psi^\dagger\rangle$. This $M$ is listed under OP. $M$ must contain an even number of Pauli matrices to because of fermion anticommutivity but may take complex values. 
\begin{longtable}{|p{.55cm}|p{2.7cm}|p{2.65cm}|p{2.2cm}|}
\hline\hline
Irr. & OP& $M^\alpha$ & Symmetry\\
\hline\hline
$A_1$ & $\Delta_{000}$ &
$\openone^{AB}\openone^{KK'}\tau^N_{x,y} \openone^s$ 
& $\vec{S}, C, t, R_C,\linebreak R_t, T\quad $ \\
\hline\hline
$A_2$ & $\left(\Delta_{0zx},\Delta_{0zy},\Delta_{0zz}\right)$ &
$\openone^{AB}\tau_z^{KK'}\tau^N_{x,y} \sigma_{x,y,z}$ 
& $ C, t, R_C$ \\
\hline
$A_{2,1}$
	& $ \left(0,0,1\right)$ && $S_z, T$\\ 
\hline
$A_{2,2}$	& $ \left(1,i,0\right)$& & $S_z + g$\\
\hline\hline
$B_1$ & $\left(\Delta_{z0x},\Delta_{z0y},\Delta_{z0z}\right)$ &
$\tau_z^{AB}\openone^{KK'}\tau^N_{x,y} \sigma_{x,y,z}$ 
& $ C, t, R_t$ \\
\hline
$B_{1,1}$	& $ \left(0,0,1\right)$ && $S_z, T$\\ 
\hline
$B_{1,2}$	& $ \left(1,i,0\right)$ && $S_z+ g$\\
\hline\hline
$B_2$ & $\Delta_{zz0}$ & 
$\tau_z^{AB}\tau_z^{KK'}\tau^N_{x,y} \openone^s$ 
&$\vec{S}, C, t,\quad\linebreak R_C\cdot R_t , T$\\
\hline\hline
$E_2$ & $\left(\Delta_{xz0},\Delta_{yz0}\right)$ &
$\tau_{x,y}^{AB}\tau_z^{KK'}\tau^N_{x,y} \openone^s$ 
& $\vec{S},t, R_C,R_t$\\
\hline
$E_{2,1}$	& $\left(1,0\right)$ & & $T$\\
\hline
$E_{2,2}$	& $\left(1,i\right) $ & & $C+g$\\
\hline\hline
$E_2''$ & $\left(\Delta_{zx0},\Delta_{zy0}\right)$ &
$\tau_{z}^{AB}\tau_{x,y}^{KK'}\tau^N_{x,y} \openone^s$
& $\vec{S},C, R_C,R_t$\\
\hline
$E_{2,1}''$	& $\left(1,0\right)$  & & $T$\\
\hline
$E_{2,2}''$	& $\left(1,i\right) $ & & $t+g$\\
\hline\hline
$E_1$ & $\left(\!\Delta_{x0x},\Delta_{x0y},\Delta_{x0z};\right.\linebreak\left.
\Delta_{y0x},\Delta_{y0y},\Delta_{y0z}\!\right)$&
$\tau_{x,y}^{AB}\openone^{KK'}\tau^N_{x,y} \sigma_{x,y,z}$ 
 & $t$\\
\hline
$E_{1,1}$	& $\left(0,0,1;0,0,0\right)$  & &$S_z, R_C, T$\\
\hline
$E_{1,2}$	& $\left(1,i,0;0,0,0\right) $  & &$S+g, R_C$\\
\hline
$E_{1,3}$	& $\left(0,0,1;0,0,i\right) $  & &$S_z,C+g, R_C\!\cdot\!T$\\
\hline
$E_{1,4}$	& $\left(\alpha, i\alpha, i\beta;i\alpha,-\alpha,\beta\right) $ & & $C + S+ 2g$\\
\hline
$E_{1,5}$	& $\left(1,i,0;i,-1,0\right) $ & & $C+ g, S + g$\\
\hline\hline
$E_1''$ & $\left(\!\Delta_{0xx},\Delta_{0xy},\Delta_{0xz};
\right.\linebreak\left.
\Delta_{0yx},\Delta_{0yy},\Delta_{0yz}\!\right)$ &
$\openone^{AB}\tau_{x,y}^{KK'}\tau^N_{x,y} \sigma_{x,y,z}$ 
& $C$\\
\hline
$E_{1,1}''$	& $\left(0,0,1;0,0,0\right) $  && $S_z, R_t, T$\\
\hline
$E_{1,2}''$	& $\left(1,i,0;0,0,0\right) $  & &$S+g, R_t$\\
\hline
$E_{1,3}''$	& $\left(0,0,1;0,0,i\right) $  & &$S_z,t+g,R_C\cdot T$\\
\hline
$E_{1,4}''$	& $\left(\alpha,i\alpha,i\beta;i\alpha,-\alpha,\beta\right) $ & & $t + S+ 2g$\\
\hline
$E_{1,5}''$	& $\left(1,i,0;i,-1,0\right) $ & & $t+ g, S + g$\\
\hline\hline
$G$ & $\left(\Delta_{xx0},\Delta_{xy0},
\right.\linebreak\left.
\Delta_{yx0},\Delta_{yy0}\right)$ &$\tau_{x,y}^{AB}\tau_{x,y}^{KK'}\tau^N_{x,y} \openone^s$ & $None$ \\
\hline
$G_1$	& $\left(1, 0, 0, 1\right)$ && $ C+t, R_C\cdot R_t, T$ \\
\hline
$G_2$	& $\left(0,0, 0 ,1\right)$  & &$ R_C, R_t, T$ \\
\hline
$G_3$	& $\left(0, 0, -i, 1\right)$  && $ C+g, R_C, R_t\cdot T$ \\
\hline
$G_4$	& $\left(0, -i, 0, 1\right)$  & &$ t+g, R_t, R_C\cdot T$  \\
\hline
$G_5$	& $\left(1, i, i, -1\right)$  & & $C+g, t+g,\quad\linebreak R_C\!\cdot\!R_t\cdot\!T\quad$\\
\hline\hline
\end{longtable}


\begin{thebibliography}{99}
\bibitem{myfootnote1}
For generic choices of the coupling constants these interactions reduce the symmetry of the model
from the artificial $SU(4)\otimes U(1)\otimes U(1)$ down to $SU(2)\otimes U(1) \otimes  U(1)\otimes \mathcal{D}_{3d}$ 
This is still larger than the original symmetry group as $\tau_{z}^{KK'}$ does not generate continuous rotations in the
original space group. The continuous $\tau_{z}^{KK'}$
rotation is broken by Umklapp terms down to $Z_3$. However, the lowest order Umklapp term is of the form
$H_{umklapp} \sim  (\psi^\dagger \tau^{KK'}_{+}\tau^{AB}_{z}\psi)^3 + h. c.$ Since this contains six fermion creation and annihilation operators it will be strongly RG irrelevant and we will not consider it.
\bibitem{myfootnote2} Equations (\ref{eq:gRG}) and (\ref{eq:DefTensors}) appeared in Ref.  ~\onlinecite{Letter} with several typos. Furthermore, 
a misdefinition of $\ell$ led to incorrect statements on the critical temperature of nematic state. The errors have been corrected in the present paper.
\end{thebibliography}
\end{document}